\documentclass[sn-standardnature]{sn-jnl} 

\usepackage{textgreek}
\usepackage{paralist}
\usepackage[separate-uncertainty=true]{siunitx}
\usepackage{xr}
\jyear{2023}%

\makeatletter

\begin{document}
	
\title[Anticrossing of a PNR mode and a single Qdot at RT]{Anticrossing of a plasmonic nanoresonator mode and a single quantum dot at room temperature}







\author[1]{\fnm{Daniel} \sur{Friedrich}}

\equalcont{These authors contributed equally to this work.}

\author*[1]{\fnm{Jin} \sur{Qin}}\email{jin.qin@uni-wuerzburg.de}
\equalcont{These authors contributed equally to this work.}

\author[1]{\fnm{Benedikt} \sur{Schurr}}
\equalcont{These authors contributed equally to this work.}

\author[2]{\fnm{Tommaso} \sur{Tufarelli}}

\author[1]{\fnm{Heiko} \sur{Groß}}

\author*[1]{\fnm{Bert} \sur{Hecht}}\email{hecht@physik.uni-wuerzburg.de}

\affil*[1]{\orgdiv{Nano-Optics and Biophotonics Group, Experimentelle Physik 5}, \orgname{Physikalisches Institut, Universit\"at W\"urzburg}, \orgaddress{\street{Am Hubland}, \city{W\"urzburg}, \postcode{D-97074}, \country{Germany}}}
\affil[2]{\orgdiv{School of Mathematical Sciences and Centre for the Mathematics and Theoretical Physics of Quantum Non-Equilibrium Systems}, \orgname{University of Nottingham}, \orgaddress{\city{Nottingham}, \postcode{NG7 2RD}, \country{United Kingdom}}}
\abstract{
Room-temperature strong coupling of a single quantum emitter and a single resonant plasmonic mode is a key resource for quantum information processing and quantum sensing at ambient conditions. To beat dephasing, ultrafast energy transfer is achieved by coupling single emitters to a plasmonic nanoresonator with an extremely small mode volume and optimal spectral overlap. Typically, normal mode splittings in luminescence spectra of single-emitter strongly-coupled systems are provided as evidence for strong coupling and to obtain rough estimates of the light-matter coupling strength $g$ \cite{leng_strong_2018,Park_tip,gupta_complex_2021}. However, a complete anticrossing of a single emitter and a cavity mode as well as the characterization of the uncoupled constituents is usually hard to achieve. Here, we exploit the light-induced oxygen-dependent blue-shift of individual CdSe/ZnS semiconductor quantum dots to tune their transition energy across the resonance of a scanning plasmonic slit resonator after characterizing both single emitter and nano resonator in their uncoupled states. Our results provide clear proof of single-emitter strong light-matter coupling at ambient condition as well as a value for the Rabi splitting at zero detuning (\SI{100}{\milli\electronvolt}), consistent with modeling, thereby opening the path towards plexitonic devices that exploit single-photon nonlinearities at ambient conditions.

}


\maketitle
	
\section{Introduction}
A single two-level system strongly coupled to a single mode of an electromagnetic field of a resonator exhibits a characteristic single-photon nonlinearity in its energy spectrum. The observed splitting of the emerging eigenmodes of the strongly-coupled system scales with the square root of the number of photons in the system. That is, the addition of a single photon to the system changes its response to a follow-up photon. Such behavior, if realized at ambient conditions, holds promise to overcome the need for cryogenics in quantum information processing and quantum sensing thus unlocking an enormous potential for applications. \cite{fink2008climbing,khitrova_vacuum_2006,delValle_quantum_2009,Bitton_plasmonics_2019}

To achieve strong coupling at ambient conditions, many studies exploited the square-root scaling of the coupling strength with the number of emitters that couple to the same cavity mode, e.g. by making use of j-aggregates or otherwise densely-packed emitter systems\cite{Schlather_nanolett,PhysRevLett.114.157401,PhysRevLett.118.237401,kleemann_strong-coupling_2017,Stuehrenberg_nl}. In such systems a coupled bright state of many emitters hybridizes with the cavity mode leading to interesting collective effects, such as enhanced photo-chemistry, enhanced conductivity and possibly light-induced superconductivity\cite{Ebbesen_science}. However, the price to pay in these systems is that the single-photon nonlinearity of the associated Tavis-Cummings model decreases with the number of emitters. To retain the single-photon nonlinearity, there has been a quest for solid evidence of reaching single-emitter strong light-matter coupling (SC) at ambient conditions\cite{Tufarelli_PhysRevResearch}. Experiments are based on the idea that the ultrasmall mode volumes of plasmonic nanoresonators should lead to strong coupling in spite of their low quality factors due to intrinsic losses and dephasing.   

The necessary small mode volume of the nanoresonator at ambient conditions requires solutions for positioning a single emitter inside the hotspot of a plasmonic nanoresonator with nanometer precision. One approach has been to use a random distribution of single emitters that are spread at low concentration on top of plasmonic nanoresonators, such as arrays of bow-tie antennas\cite{gupta_complex_2021,Bitton_plasmonics_2019,santhosh2016vacuum} or nanoparticle-on-mirror geometries\cite{chikkaraddy2016single,ojambati_quantum_2019}. Scanning-probe techniques which position a plasmonic nanoresonator with nanometer accuracy on a surface also have proven useful to achieve the necessary positioning accuracy while providing the possibility to vary the coupling strength deliberately at any time to study the uncoupled entities\cite{Park_tip}. In all experiments photoluminescence spectra (PL) that exhibit two clearly split peaks shifted red and blue with respect to the uncoupled resonance of cavity and emitter system are generally presented as proof of strong coupling. However, the observation of a single split spectrum for a fixed detuning, only, may be considered a weak proof for SC since vast room is left for alternative explanations\cite{westmoreland2019properties}. Observation of anti crossing of emitter and resonator resonance together with a characterization of the uncoupled partners represents a much stronger and well-established evidence for SC\cite{khitrova_vacuum_2006,yoshie2004vacuum}. Yet, in room temperature SC experiments so far, anti-crossing data has either been missing, or obtained by stitching measurements from different coupled systems, which clearly falls behind the ideal single emitter - single resonator experiment\cite{gros_near-field_2018,chikkaraddy2016single,li2022room,ojambati_quantum_2019}. Besides proving strong coupling, anti-crossing curves are also necessary to quantify the coupling strength at zero detuning because a possible nonzero detuning between the cavity and two-level system will increase the observed splitting of the spectra thus mimicking a larger coupling strength. Furthermore, in all single-emitter strong coupling experiments reported so far only a statistical characterization of emitter and resonator is performed, leaving considerable uncertainties as of the properties of the individual systems and in particular hampering the exclusion of detuning between emitter and resonator. Therefore, only qualitative estimates of the coupling strength are obtained and the possibility of observing uncoupled detuned spectral peaks cannot be fully excluded.  

Here we demonstrate complete anticrossing of the PL peak of a single CdSe/ZnS semiconductor quantum dot (Qdot) and the second-order resonance of a scanning plasmonic nanoslit resonantor at room temperature, which have both been characterized separately before the coupling experiment (Fig.~\ref{fig1_setup}). Tuning of the Qdot PL is achieved using a light-induced spectral shift in presence of oxygen. The obtained anticrossing resonances allow us to extract the light-matter coupling strength at zero detuning with excellent precision. The repeatability and the high degree of control of the experiment opens the road towards deterministic fabrication of plexitonic devices at ambient conditions providing single-photon nonlinearities.


\section{Results \& Discussion}

Our experiments are based on plasmonic slit nanoresonators (PNR) fabricated at a corner of a mono-crystalline gold microplatelet\cite{gros_near-field_2018}. 
Such PNRs are used as scanning probe tips in an atomic force microscope (AFM) (Fig.~\ref{fig1_setup}) in a setup that combines an AFM with a scanning confocal optical  microscope (see supplementary Fig.~\ref{SI:setup}). Making use of the atomic-force-microscopy capabilities of the setup provides nanometer precision in positioning Qdots beneath the PNR tip while the confocal microscope records the luminescence of the system. It is possible to probe different Qdots using the same PNR as well as to record the resonances of uncoupled PNRs and Qdot PL spectra before and after the coupling experiment. Changing the relative position of a PNR and a single Qdot furthermore allows us to probe the distance-dependent coupling strength (see Fig.~\ref{fig1_setup}a).

We use He-ion beam milling to fabricate PNRs with suitable resonances. A scanning electron microscopy (SEM) image of a fabricated PNR probe tip is displayed in Fig.~\ref{fig1_setup}b. The second-order plasmonic resonance of the \SI{210}{\nano\meter} long and \SI{15}{\nano\meter} wide PNR is found at \SI{1.916}{\electronvolt} (\SI{647}{\nano\meter}, Q-factor: \si{20}) by distinct shaping of the recorded gold PL and subsequent fitting (see supplementary Fig~\ref{fig4_PNR}d) as confirmed by finite-difference time-domain (FDTD) simulations. The resonance frequency can be controlled by tuning the length and with of the PNR\cite{gros_near-field_2018}. Details of the PNR fabrication and the characterization procedure are described in the supplementary information~\ref{SI:plasmonic nanoresonator} as well as in \cite{gros_near-field_2018}.

We investigate the coupling of PNRs to single colloidal CdSe/ZnS semiconductor nanocrystals (Thermo Fisher Scientific Inc., QDot655 ITK Q21321MP) with an average emission energy of \SI{1.893}{\electronvolt} (\SI{655}{\nano\meter}, Q-factor: \si{34}) dispersed on a glass cover slip at a coverage of about 1 emitter/$\si{\micro\meter\squared}$ by spin coating. In a second step, the cover slip is spin-coated with \SI{10}{\nano\meter} polymethylmethacrylate (PMMA) film to immobilize the Qdots.
An atomic force microscopy (AFM) image of such a sample is shown in supplementary Fig.~\ref{fig4_QDots} a,b. When analyzing the $2^{\mathrm{nd}}$-order autocorrelation function, $g^2(\tau)$, of the Qdots' emitted PL, a $g^2(0)=0.08$ is typically obtained (Fig.~\ref{fig1_setup}c) at zero time delay, $\tau$, alongside the typical blinking behavior. Both observations signify the fact that a single emitter is investigated. Further details of the single emitter characterization are presented in the supplementary information~\ref{SI:quantum dot}.

\begin{figure*}[htp!]
   \centering
    \includegraphics[width=1\linewidth]{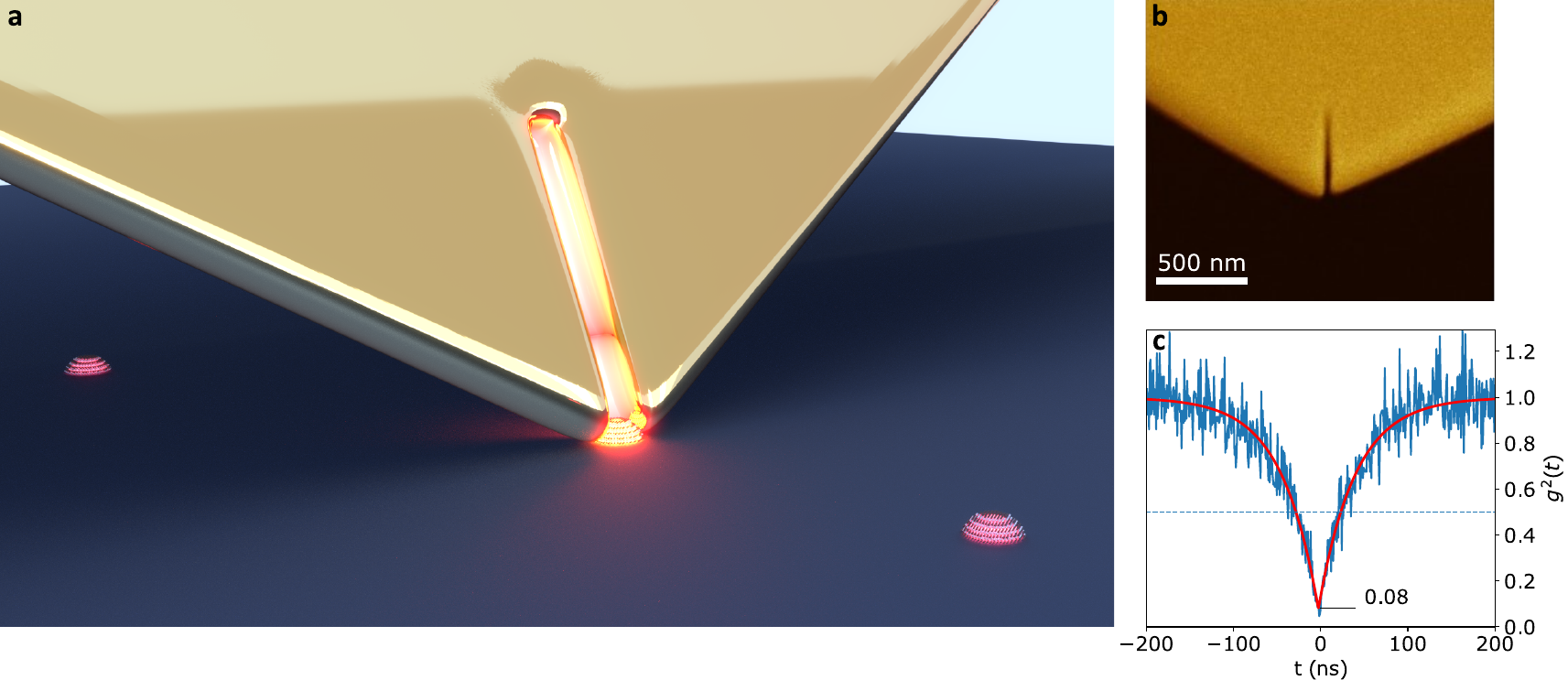}
    \caption{\textbf{Setup.} - \textbf{a} A gold microplatelet scanning plasmonic nanoresonator (PNR) in close contact to a quantum dot (artistic representation). \textbf{b} SEM image of a fabricated PNR.  \textbf{c} Typical $2^{\mathrm{nd}}$-order autocorrelation function, $g^2(\tau)$, recorded for a single QD.
    }
    \label{fig1_setup}
\end{figure*}


After pre-characterizing the PNR and the Qdot individually to ensure spectral overlap as well as single-emitter character of the Qdot emission, the PNR is positioned in close proximity to the Qdot to investigate possible strong light-matter interaction (see \ref{SI:setup}). To this end, a $\lambda_\mathrm{exc} = \SI{532}{\nano\meter}$ laser spot (power: $<\SI{1}{\micro\watt}$, intensity: \SI{1.7e7}{\watt/m^2}) is focused at the PNR tip and then the Qdot is moved beneath the tip and scanned. Note that at the low excitation intensity that we are utilising, gold PL is too weak to be detected above background. However, at the selected wavelength, the light acts as an efficient non resonant pump of the Qdots which results in PL at around \SI{655}{\nano\meter} which then couples with the matching PNR resonance (see supplementary~\ref{SI:plasmonic nanoresonator}). It is important to reiterate that the detected PL spectra in our measurements, due to the low excitation intensity used, can either show: (i) spectra of
uncoupled Qdots or (ii) spectra of the coupled system. A mere combination of uncoupled Qdot and PNR can never be observed, since the emission of the latter would be too weak in comparison to the Qdot.\\

\begin{figure*}[htp!]
   \centering
    \includegraphics[width=1\textwidth]{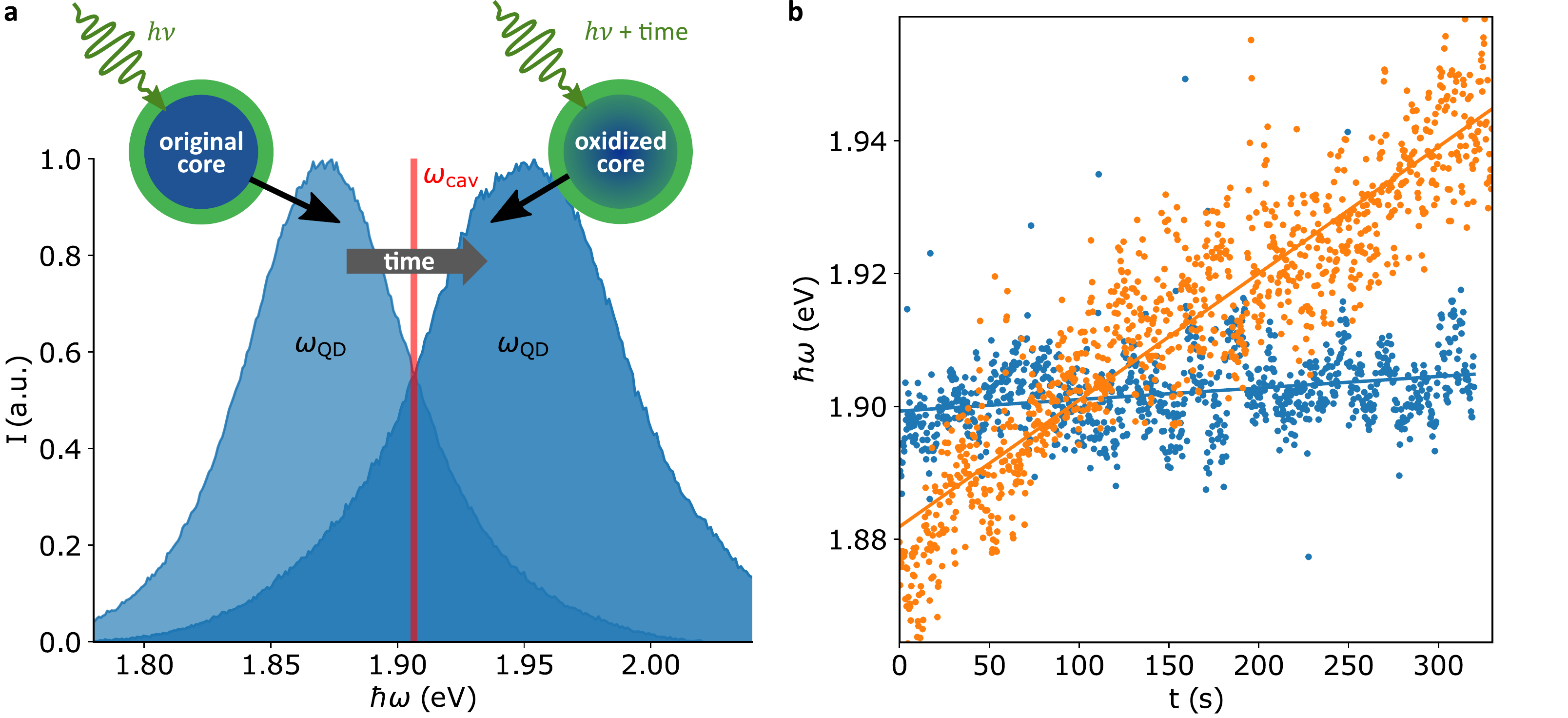}
    \caption{\textbf{Qdot blue-shift.} \textbf{a} Illustration of the core size reduction due to oxidation and the resulting increased transition energy. \textbf{b} Comparison of the resonance peaks fitted with a Lorentzian for an individual Qdot at ambient conditions in air (orange) and in argon atmosphere (blue).}
    \label{fig_QD-blue-shift}
\end{figure*}

To achieve a complete anticrossing of PNR and QDot, we make use of the well-known light-induced oxygen-dependent blue-shift of QDots\cite{nirmal_fluorescence_1996,van_sark_blueing_2002}. We confirmed this effect for the used Qdots and found that it can indeed be prevented in an oxygen-free atmosphere, e.g.~in argon, and in absence of the pump light (see Fig.~\ref{fig_QD-blue-shift}). As a possible explanation for the observed pronounced light-induced blueshift in presence of oxygen, oxidation of the outer shell of the core has been put forward, which effectively shrinks the core leading to a controlled spectral shift of the Qdot PL by up to \SI{100}{\milli\electronvolt} over several minutes of continuous illumination\cite{van_sark_blueing_2002}. To obtain anticrossing we start out with a Qdot that is red-detuned with respect to the bare-cavity resonance. During the experiment, the Qdot resonance will then sweep over the cavity resonance as function of time to end up in a blue-detuned state. By monitoring the spectra of the emitted PL during the whole duration of the blue shifting process, an anti-crossing of Qdot PL and PNR resonance is recorded. The linear blue-shift of a bare Qdot as a function of time in air, respectively its absence in argon, are illustrated in Fig.~\ref{fig_QD-blue-shift}b for a representative Qdot in air and in argon-atmosphere. \\

Within about 300s, the Qdot in air shows a blue-shift of about \SI{62.8}{\milli\electronvolt} compared to its original resonance, while in argon atmosphere for the same Qdot the resonance energy remains unchanged. The blue-shifting of the Qdot resonance is an irreversible process that continues to progress with longer measurement times and typically ends with the final photo-bleaching of the Qdot\cite{van_sark_blueing_2002}. \\





\begin{figure*}[htp!]
   \centering
    \includegraphics[width=.99\linewidth]{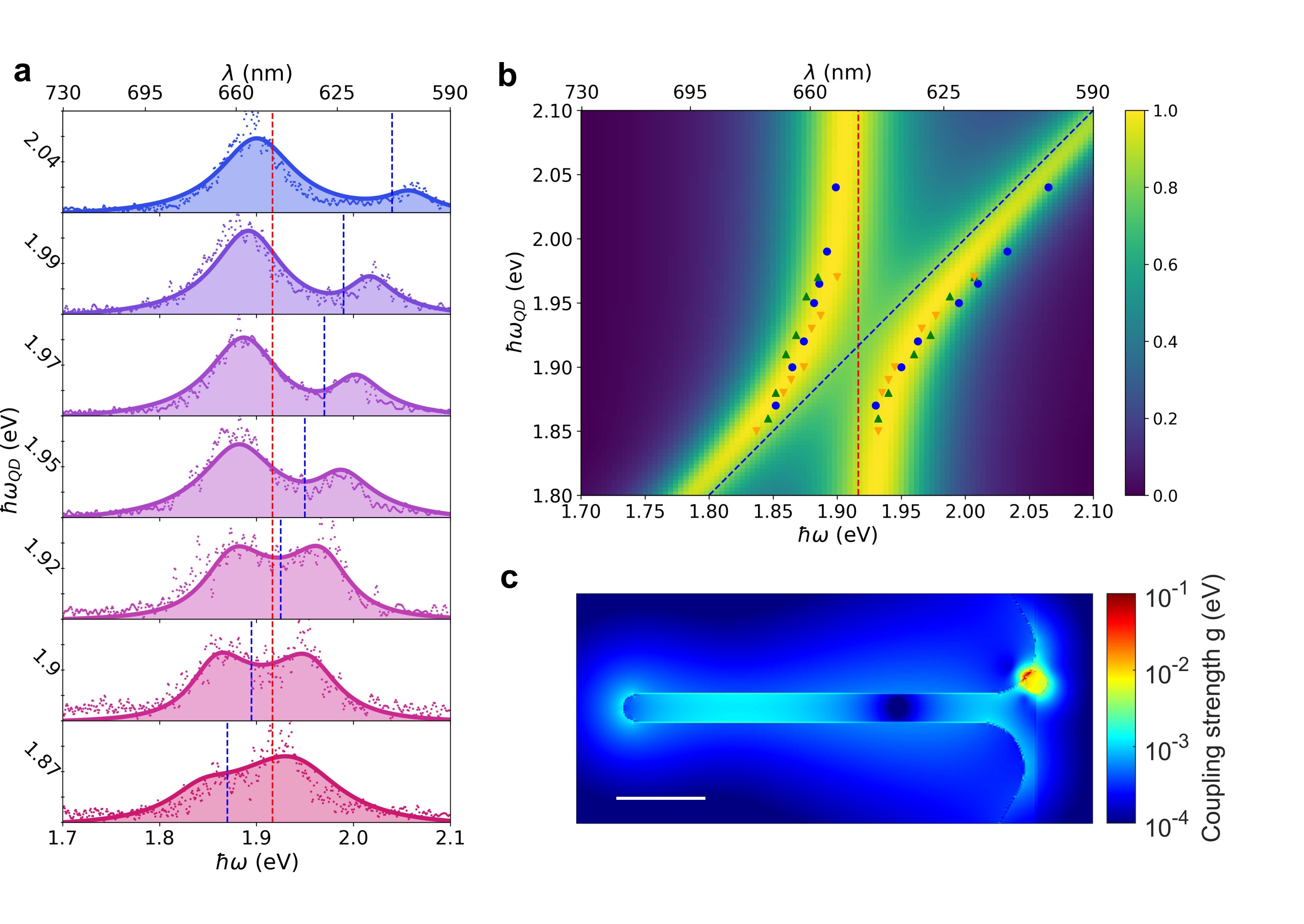}
    \caption{\textbf{Anticrossing.}  \textbf{a} PL (tiny dots) time series (time sequence: from bottom to up) with fits (solid line) resulting from the JC model. Red and blue dashed lines are indicating the cavity and the fitted Qdot resonance, respectively. \textbf{b} Background: Anti-crossing of a fixed cavity resonance ($\omega_\mathrm{a}$, dashed red line) and a linearly varying Qdot resonance ($\omega_{\sigma}$, dashed blue line) calculated by the quantum model. Peak positions of three SC experiments (with three different Qdots) are labeled by different shapes and colors. All three Qdot exhibit a complete and consistent anticrossing. The parameters of the underlying modelled anticrossing curve can be found in the supplementary information~\ref{SI:quantum_model}. \textbf{c} Map of the coupling strength $g$ in \si{\electronvolt} (log scale) as obtained by classical electromagnetic modelling (FDTD) using a point-like dipole momentum of \SI{5}{Debye}. The map is rotated $90^{\circ}$ counterclockwise compared with the inset in Fig.\ref{fig_FDTD_simulationarea}. Scale bar: \SI{50}{\nano\meter}.} 
    \label{fig3_anticrossing}
\end{figure*}

To perform an anticrossing experiment we pick a Qdot with a sufficiently red-detuned resonance compared to the bare resonance of the PNR taking advantage of the inhomogeneous broadening of the Qdot ensemble and the possibility to tune the PNR resonance at will. We then continuously record PL spectra at a \SI{33}{\milli\second} time resolution. 
The resulting photoluminescence time series (from bottom to top) of a strongly coupled Qdot-PNR system is shown in Fig.~\ref{fig3_anticrossing}a.
The only changing parameter in these spectra is the time-dependent, oxygen-induced blue-shift of the Qdot emission. The detailed fitting strategy is described in the supplementary material~\ref{SI:quantum_model}. The spectra displayed in Fig.~\ref{fig3_anticrossing}a are selected around equidistant time intervals from a continuous measurement to obtain an approximately uniform energy spacing (see supplementary Fig.~\ref{fig_spectra_selection}). The dashed lines indicate the measured cavity bare resonance (red) and the fitted Qdot resonance (blue) which indeed shows the expected sweeping of the Qdot resonance over the cavity resonance leading to the observed anticrossing. \\

As a function of time, the strongly coupled spectra change from red-detuning to blue-detuning. The resulting spectra generally exhibit two peaks and can be nicely modeled by a quantum Jaynes-Cummings model\cite{bruce_jaynes-cummings_1993}. This model is widely used for describing the light-matter interaction of a single two-level system and a single-mode cavity. In order to model incoherent pumping, radiative decay (dissipation) and pure dephasing, we model the system via a master equation in Lindblad form. The validity of this simplified description is sometimes questioned due to a possible mode overlap of neighboring modes\cite{franke_impact_2022}. However, for the present slit design the crosstalk of adjacent modes is highly suppressed\cite{hensen_spatial_2019} which is also verified by our characterization of the uncoupled PNR via gold PL showing only one resonance order within the spectral range relevant for coupling (see supplementary information \ref{SI:plasmonic nanoresonator}). Furthermore, the PNR slit modes exhibit a quadrupolar character leading to enhanced Q factors around 20 mostly due to reduced radiative losses. It is therefore safe to assume that our PNR indeed exhibits only a single plasmonic mode in the coupling range. \\

An important conclusion from our observations is that after sufficiently long coupling experiments with Qdots the resulting SC spectra will typically always show a blue-detuned Rabi-splitting. This doesn't mean that there is no strong coupling anymore. The observation of two peaks in the PL can still only be explained in this way. Yet, the observed splitting is far larger than the splitting at zero detuning and can lead to a significant over-estimation of the coupling strength. Indeed several reports demonstrated blue-shifted strong-coupling spectra of individual Qdots to plasmonic resonators\cite{hartsfield_single_2015, leng_strong_2018, gupta_complex_2021}. The reported spectra are difficult to interpret quantitatively since the bare resonances of single quantum dot and plasmonic cavity are difficult to estimate and a complete anticrossing is not available. \\


Time dependent measurements of three different Qdots, but with the same PNR, are collected in Fig~\ref{fig3_anticrossing}b. The two maxima of the two peaks, $\omega_{-}, \omega_{+}$, of the recorded Rabi-splitted spectra are indicated with a dot for each fitted peak of the quantum model. A typical anticrossing map predicted by the quantum model with a fixed cavity resonance (red dashed line) and linear detuned Qdot resonance (blue dashed line) is displayed in the background. Note that each individual Qdot produces a complete anticrossing with the PNR. All plots are in excellent agreement with one another and can be well modelled using the same parameters which allows us to extract the effective coupling rate $g$ at zero detuning from the quantum model which amounts to about \SI{50}{\milli\electronvolt}. \\

For a strongly coupled system, at least one complete Rabi oscillation is needed in the time domain, which requires $2g>(\gamma_\mathrm{a}+\gamma_\mathrm{QD})/{2}$ where $\gamma_{a}$ and $\gamma_\mathrm{QD}$ are total decay rates of the PNR and quantum dot, respectively. Both parameters are extracted by fitting the FWHM of the PL curves of both uncoupled entities. In our case, the PNR has a high Q value of about 20 because of its quadruple mode pattern. This results in $\gamma_{\mathrm{a}} = \SI{95}{\milli\electronvolt}$ while $\gamma_\mathrm{QD}$ is roughly about \SI{56}{\milli\electronvolt}. Our hybrid system therefore fulfils the above criterion due to the low-loss plasmonic cavity and the high single-emitter coupling strength. \\

Second order correlation, $g^2(\tau)$, measurement have also been performed with the system being strongly coupled. However, in contrast to the clear dip below 0.5 at time-delay $\tau=0$, observed for the uncoupled Qdots, we couldn't detect such a dip in strong coupling. Indeed, when the PNR is strongly coupled to the Qdot, the decay rate of the hybrid system is $(\gamma_{a}+\gamma_\mathrm{QD})/2$, which is in the fs-range rendering a dip at zero delay impossible to resolve since the time bin resolution in our experiment is limited to \SI{0.1}{\nano\second}. The absence of a correlation dip in strong coupling is therefore consistent with strong coupling in our system. \\

In the dipole approximation\cite{torma_stong_2014} the coupling strength $g(\boldsymbol{r})$ can also be estimated as   $g(\boldsymbol{r})=\boldsymbol{\mu}\sqrt{\frac{\hbar\omega}{2\epsilon_0V(\boldsymbol{r})}}$ , where $\boldsymbol{\mu}$ is the dipole moment of the quantum dot and $V(\boldsymbol{r})$ is the effective mode volume of the PNR's resonant mode at position $\boldsymbol{r}$. FDTD simulations have been performed to accurately determine the mode volume based on quasi-normal-mode theory\cite{sauvan_theory_2013}. Also, the finite size of quantum dot is considered by modelling it as a sphere with Lorenzian permittivity (see supplementary information~\ref{SI:classical_model}) to mimic its dipolar two-level transition\cite{Schlather_nanolett}. On the basis of uncoupled Qdot lifetime measurement, a dipole moment of \SI{5}{Debye} is assigned to the Qdot in the simulated coupling-strength map displayed in Fig.~\ref{fig3_anticrossing}c. Inspection shows that the coupling strength $g$ indeed reaches up to \SI{50}{\milli\electronvolt} in agreement with our experimental results when the PNR is in close proximity to the Qdot. The variance of the coupling-strength map also indicates the fact that near-field local effects greatly modify the coupling behaviour, especially in the close proximity regime\cite{gros_near-field_2018}. 
It is also possible to check the presence of strong coupling in our system by performing both classical time and frequency domain FDTD simulations. As expected intuitively, the spectral splitting as well as the corresponding energy exchange in the time domain between the PNR and the Qdot can be observed (See supplementary information~\ref{SI:classical_model}) . 
Both of these studies reveal a coupling strength compatible with our experiments and the quantum model.   


\section{Conclusion and Outlook}
In summary, we have experimentally demonstrated strong coupling of a PNR scanning probe coupled to a single Qdot at ambient conditions. Our findings provide compelling evidence of strong coupling, as we not only observe the typical energy splitting in the spectral domain but also a complete anticrossing map in PL spectra. This was achieved by utilizing the light-induced oxygen-dependent blue-shift in core/shell Qdots while keeping the resonance of the PNR constant. Our scanning approach also enabled us to carefully check each uncoupled partner before and after the coupling experiment. Our experimental results are consistent with the Jaynes-Cummings model, which allows us to extract the coupling spectra and determine a mean coupling energy of \SI{50}{\milli\electronvolt} by fitting experimental spectra, and with classical field simulations as to be expected in the single-excitation regime. Our findings take us another step forward towards using strong light-matter coupling as a resource in quantum information and quantum sensing schemes as well as towards exploiting the single-photon nonlinearity of the underlying Jaynes-Cummings model even at room temperature.

\section{Author contributions}
B.H. supervised and initiated the experiments. D.F., B.S. and H.G. fabricated the plasmonic nanoresonators and performed SEM characterization. D.F. and B.S. set up and optimized the AFM setup. B.S. characterised the properties of the quantum dots. D.F., B.S and J.Q. performed SC measurements. J.Q. implemented the FDTD simulations. J.Q. and T.T. developed quantum dynamical simulations. D.F. and J.Q. processed the SC spectra. J.Q., D.F., B.S. and B.H. wrote the manuscript with input from all co-authors. All authors contributed to the discussion and have given approval to the final version of the manuscript.

\section{Acknowledgement}
This project was partly funded within the QuantERA II Programme that has received funding from the European Union’s Horizon 2020 research and innovation programme under Grant Agreement No 101017733, and with DFG (HE5618/12-1), SFI (Ireland), and NCN (Poland). We gratefully acknowledge funding by the Deutsche Forschungsgemeinschaft (DFG, German Research Foundation) under Germany’s Excellence Strategy through the W\"urzburg-Dresden Cluster of Excellence on Complexity and Topology in Quantum Matter, ct.qmat (EXC 2147, Project ID ST0462019) as well as through a DFG project (INST 93/959-1 FUGG), a regular project (HE5618/10-1), and a Reinhard-Koselleck project (HE5618/6-1). We further acknowledge funding by the "Bayerische Staatsministerium für Wissenschaft und Kunst", via the program "Grundlagen-orientierte Leuchtturmprojekte für Forschung, Entwicklung und Anwendungen im Bereich Quantenwissenschaften und Quantentechnologien" within the "Munich Quantum Valley" (IQ-Sense). We thank Ren\'e Kullock and Jingwei Zhou for helpful discussions, Enno Schatz for support with mono-crystalline gold microplatelet growth, Monika Emmerling for help with gold flake transfer, Jessica Meier for support with operating the HIM, and "qutools" for help with the HBT experiments.


\section{Supplementary information}
\subsection{Characterization}

\subsubsection{Setup}
\label{SI:setup}
\begin{figure*}[htp!]
   \centering
    \includegraphics[width=.99\linewidth]{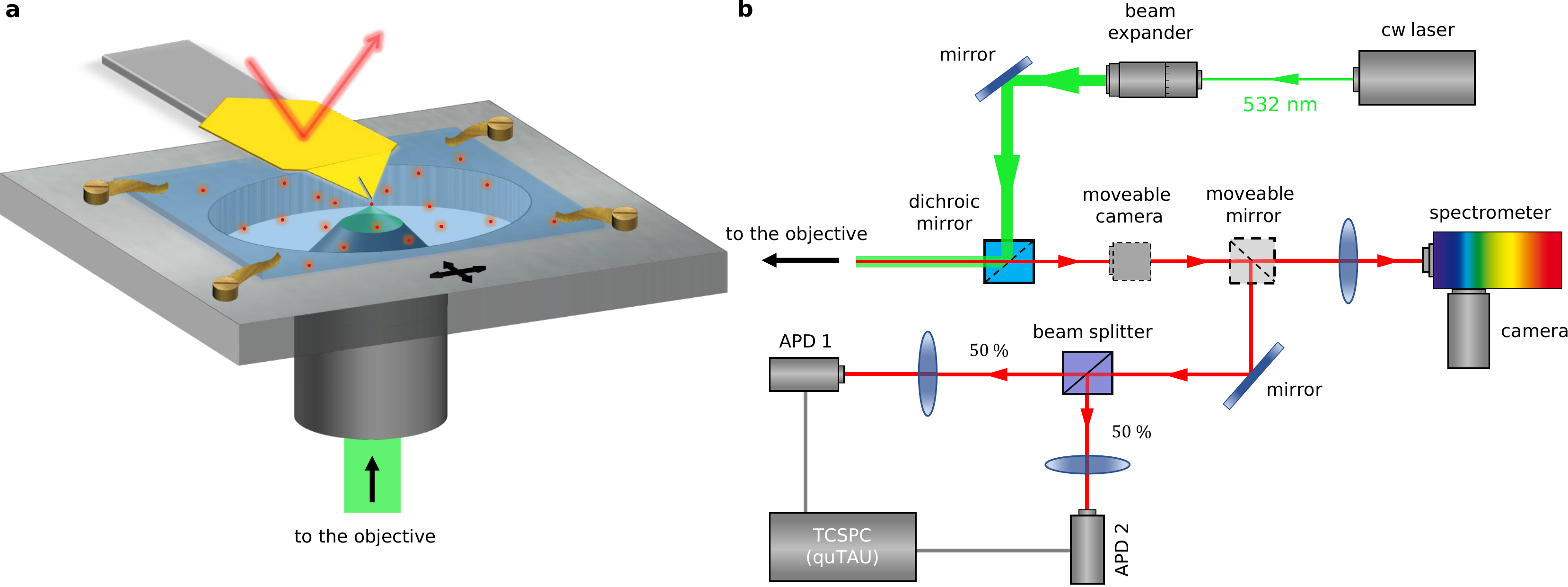}
    \caption{\textbf{Setup for strong coupling measurements}. A $532\,\mathrm{nm}$ continuous wave laser (green lines) is sent through a beam expander onto a dichroic mirror and into a microscope objective which focuses the laser light onto the sample surface and excites either the quantum dots alone or the hybrid system of emitters and cavity. The emitted signal (red lines) is collected by the microscope objective and transmitted towards the dichroic mirror (Semrock HC BS R532 1 lambda PV flat). A moveable camera allows us to observe the focal plane and the approach of the slit-resonator scanning probe towards the sample. After passing the dichroic mirror, a removable mirror guides the signal either into the spectrometer or towards two APDs. The spectrometer allows a spectral analysis of the emitted signal, whereas the APDs provide temporal information about photon arrival times which is useful to calculate photon statistics. To this end the two APDs are connected to a photon-analysis counter box (quTau Time-to-Digital Converter) which is controlled and read out by a computer.}
    \label{fig_SI_setup}
\end{figure*}

Photoluminescence (PL) measurements are performed on an inverted confocal microscope equipped with an \SI{532}{\nano\meter} continuous wave laser (AIST-NT ROU006, max. power 40\,mW), a high-numerical-aperture objective (Nikon CFI P-Apo 100x Lambda oil/ NA 1.45/ WD 0.13) as well as a spectrometer (HORIBA iHR320)/camera (Andor Newton 970p EMCCD) combination. Photon statistics are recorded with two avalanche photo diodes (APD, SPCM-AQR, APD1: SPCM-AQR-13, APD2:SPCM-AQR-14). The photon arrival events are counted by a field programmable gate array (FPGA) (qutools quTAU H+).

\subsubsection{Plasmonic nanoresonator}
\label{SI:plasmonic nanoresonator}
\begin{figure*}[htp!]
   \centering
    \includegraphics[width=.99\linewidth]{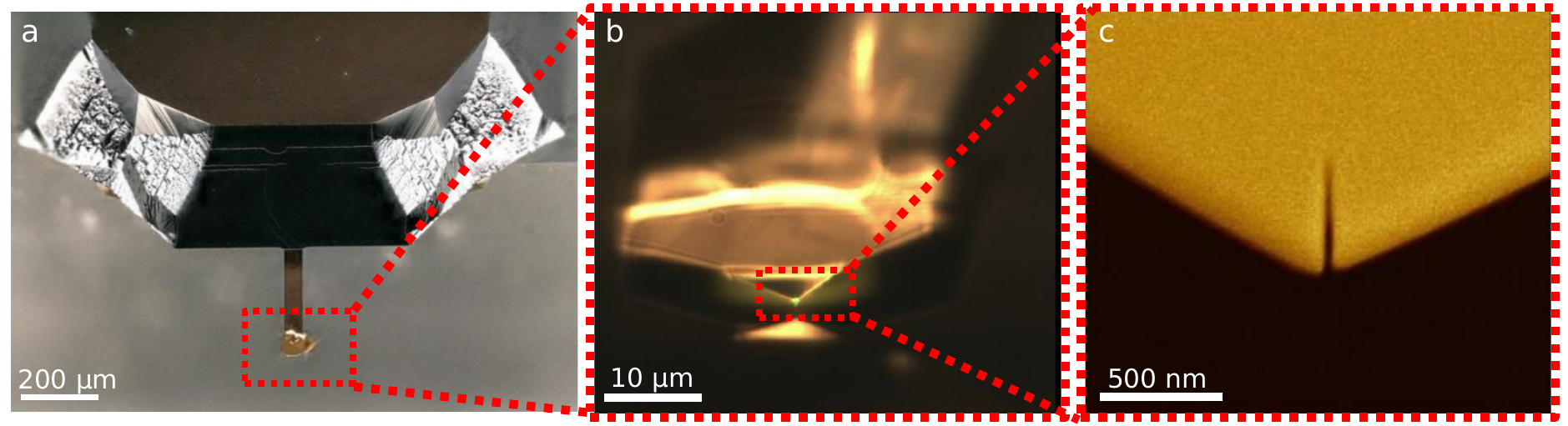}
    \caption{\textbf{Plasmonic nanoresonator}. \textbf{a} and \textbf{b} optical microscope image of the fabricated PNR on top of an AFM cantilever. \textbf{c} SEM image of the PNR with a \SI{300}{\nano\meter} long and \SI{15}{\nano\meter} wide slit. \textbf{d} Photoluminescence measurement of the PNR fitted by the sum (red solid line) of an exponential decay (dashed red line) to model the gold-PL background and a Lorentzian peak (green solid line) to the measured data (blue solid line).
    }
    \label{fig4_PNR}
\end{figure*}

 To fabricate the PNR we are using a gold microplatelet synthesized in solution based on the recipe described in \cite{huang_atomically_2010, wu_single-crystalline_2015, krauss_controlled_2018}. The microplatelet exhibits a characteristic extension of about \SI{60}{\micro\meter} and a thickness of around \SI{60}{\nano\meter}. The platelet is transferred on the top end of a contact mode AFM cantilever (doped silicon and no reflective coating, CONT-50, NanoWorld Pointprobe, NanoAndMore GmbH) with one corner extending \SIrange{5}{10}{\micro\meter} beyond the cantilever (see Fig.~\ref{fig4_PNR}a). After fabricating the plasmonic nanoresonator slit by means of a helium ion microscope, the corner is bend down by low-dose ion irradiation beneath the corner to induce a folding (see Fig.~\ref{fig4_PNR}b and \cite{gros_near-field_2018}. A scanning electron microscopy image of the fabricated PNR is displayed in Fig.~\ref{fig4_PNR}c.
The resonance of the PNR can easily be tuned to match the Qdot resonance of \SI{650}{\nano\meter} (see~\ref{SI:QD blue-shift} by adjusting the slit's length and width. The design parameters are optimized by finite-difference time-domain (FDTD Solutions, Lumerical) simulations and suggest that a \SI{210}{\nano\meter} long and \SI{15}{\nano\meter} wide slit should match the Qdot emission at \SI{650}{\nano\meter}.

\begin{figure*}[tp!]
   \centering
    \includegraphics[width=.99\linewidth]{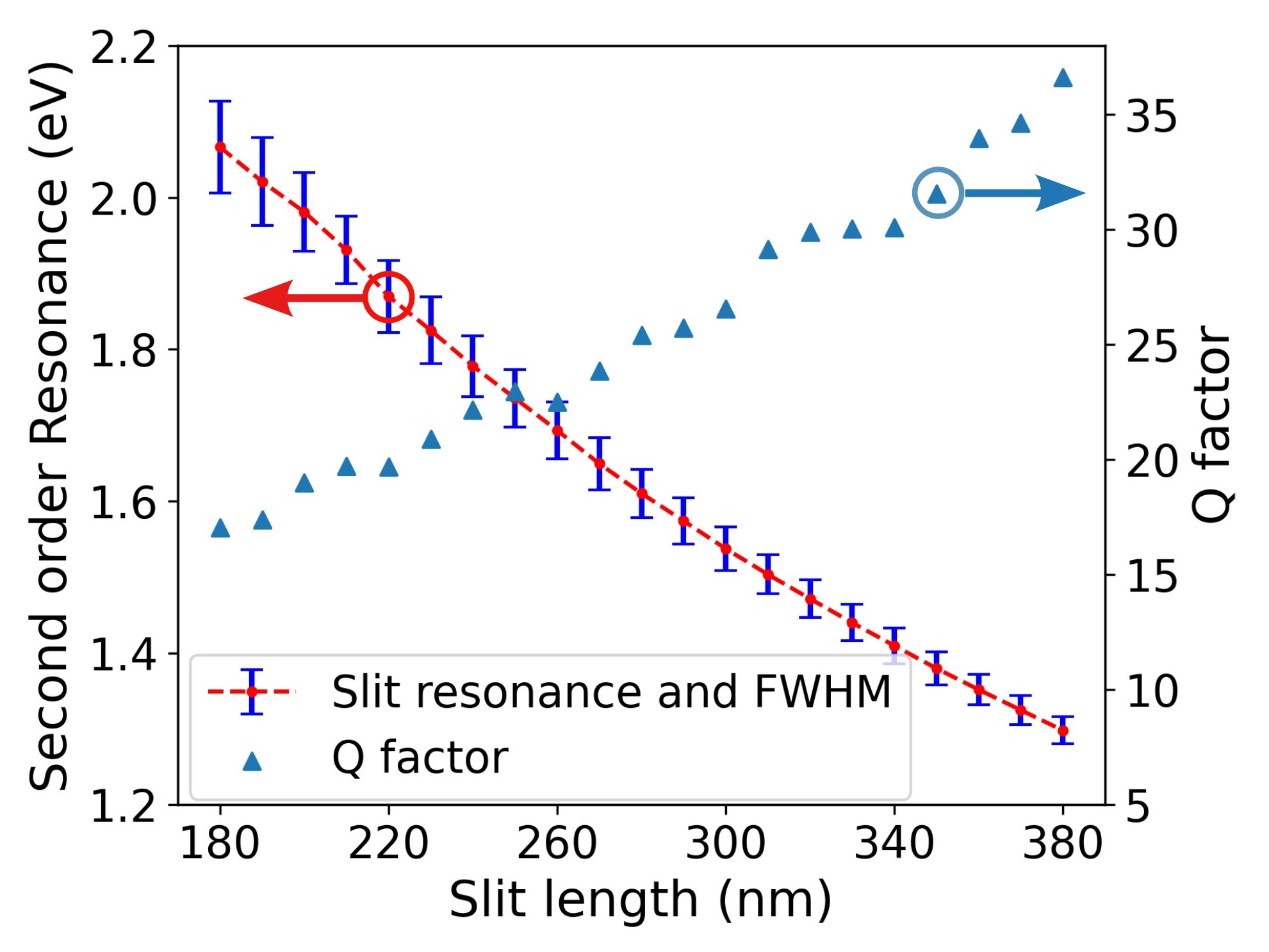}
    \caption{Slit-length dependence of the second order resonance of a PNR (red dashed line) covering slit lengths from \SI{180}{\nano\meter} to \SI{380}{\nano\meter}. The error bars indicate the full width at half maximum (FWHM) of the resonances. The blue triangles indicate the corresponding Q factors (right axis).}
    \label{fig4_cavity_Q_sim}
\end{figure*}

FDTD simulations also yield the Q-factor of the PNR resonances. To this end, a broadband electric dipole source is placed at the gap center of the PNR to excite modes of different orders. As we are interested in the second-order resonance, a dipole position along the PNR slit is chosen close to the expected antinode of the mode profile. Far-field emission is collected by recording the emitted power. Suitable temporal apodization is used to suppress dipole source contributions in the mode's field profile. Due to the quadrupolar character of the second-order mode, which minimizes radiative losses, a Q factor of \SI{20} is extracted from fitting the resulting far-field spectrum with a Lorentzian, supporting our experimental findings reported in the main text. The resonance of the PNR can be tuned by changing the slit width and length. Typically, with a larger slit length and width, the second-order resonance of the PNR red-shifts and broadens. In Fig.~\ref{fig4_cavity_Q_sim}, the changes in resonance and FWHM of the $2^{nd}$ mode are plotted as a function of the slit length ranging from \SI{180}{\nano\meter} to \SI{380}{\nano\meter} at a fixed slit width of \SI{15}{\nano\meter}. Due to the quadrupolar mode pattern a high Q factor is always maintained. 

The spectrum of the PNR is measured via the shaping of the intrinsic linear photoluminescence (PL) of gold exited by a \SI{532}{\nano\meter} continuous wave laser diode (\SI{2.2e9}{\watt/m^2}) in the vicinity of the PNR. Generally, the resonance and Q factor of the PNR are extracted from the gold PL spectrum by using a cumulative fit function witch contains an exponential decay and one Lorentzian for the cavity emission peak:
\begin{equation}
    f(\omega) = \frac{A_\mathrm{a} \mathrm{e}^{-b\omega}}{\gamma_\mathrm{a}(1+(4(\omega-\omega_{\mathrm{a}})^2/{\gamma_\mathrm{a}}^2)} + A\mathrm{e}^{-b\omega} + c  \label{eq:Pl_fitting-1}\; ,
\end{equation}
where $A_{\mathrm{a}}$, $\gamma_{\mathrm{a}}$, $\omega_{\mathrm{a}}$ indicate the amplitude, FWHM, and resonance of the PNR, respectively. The resulting spectrum is shown in Fig.~\ref{fig4_PNR}a-d for the PNR probe used in the reported experiments. 

\begin{figure*}[tp!]
   \centering
    \includegraphics[width=.99\linewidth]{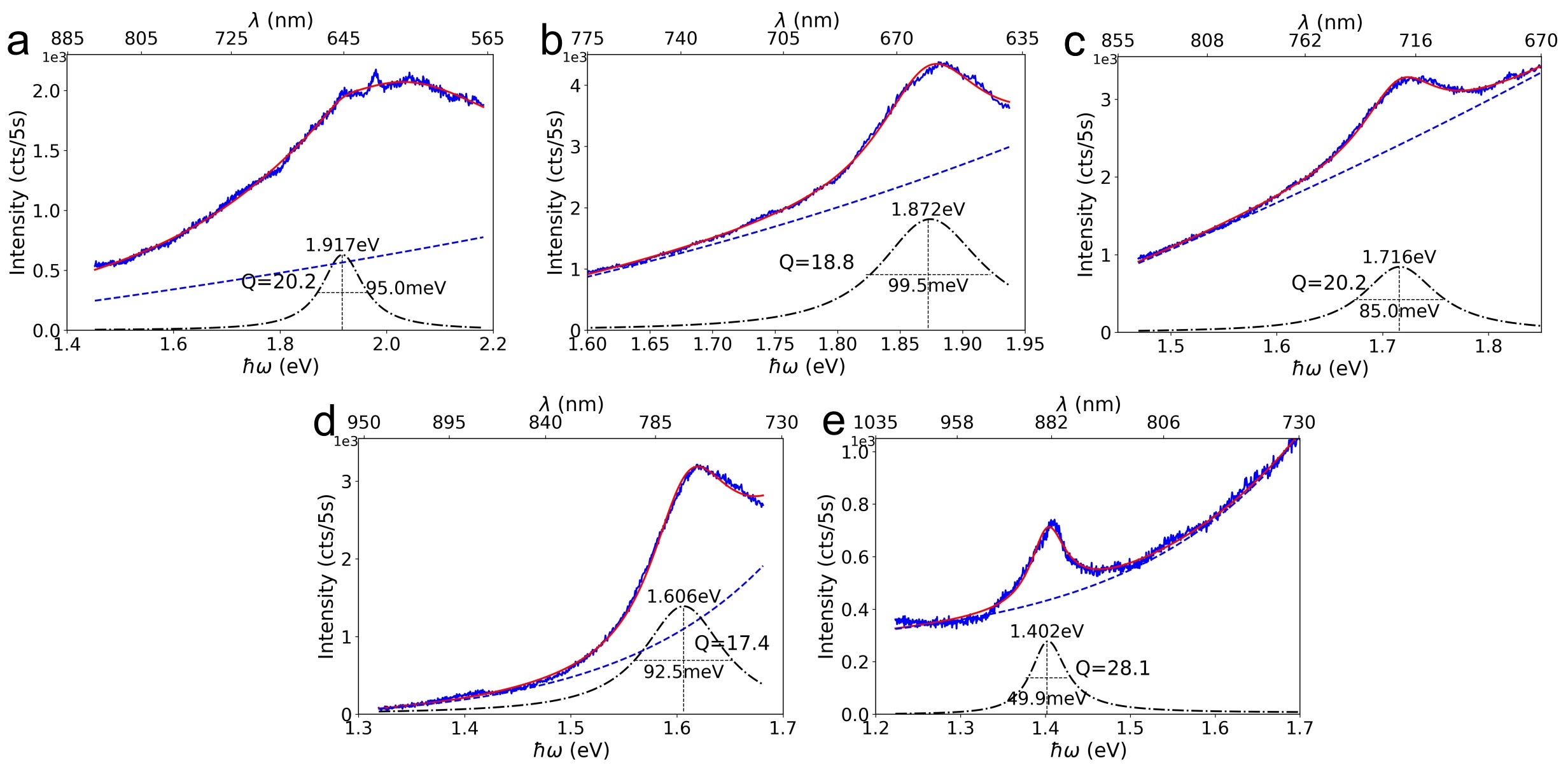}
    \caption{\textbf{Characterization of plasmonic nanoresonators with different lengths via gold PL.} The blue solid lines represent the measured PL spectra which clearly show shaping of the gold PL background due to the presence of the resonator for different slit lenghts. Spectra in \textbf{b-e} are fitted with a Lorentzian on top of a background that decays exponentially towards lower energy (red line) \eqref{eq:Pl_fitting-1}. Fitting of spectrum in \textbf{a} is done by \eqref{eq: PL_fitting-2} to add X-symmetry-point radiation into consideration. All PNRs exhibit a consistently high Q factor value.}
    \label{cavity_Q_data}
\end{figure*}
Experimental PL spectra of different slit lengths (different resonances) are displayed in Fig.~\ref{cavity_Q_data}(b-e) together with the corresponding fits. The expected slit-length dependence of the resonance is faithfully recovered as are the comparatively high Q factors. 
However, for the case of the PNR probe used for the experiments presented in the main manuscript the resonance of the PNR overlaps with interband transitions near the X symmetry point of the first Brillouin zone of monocrystalline gold\cite{imura_plasmon_2004}, which peaks roughly at \SI{650}{\nano\meter} (\SI{1.91}{\electronvolt}). This broadband transition partly obstructs the fitting of the PNR PL spectra, especially the determination of Q factor. To take interband transitions into consideration, another Lorentzian function is added to the fitting equation, which then reads as:
\begin{equation}
    f(\omega) = \frac{A_\mathrm{a} \mathrm{e}^{-b\omega}}{\gamma_\mathrm{a}(1+(4(\omega-\omega_\mathrm{a})^2/{\gamma_\mathrm{a}}^2)} +\frac{A_\mathrm{x} \mathrm{e}^{-b\omega}}{\gamma_\mathrm{x}(1+(4(\omega-\omega_\mathrm{x})^2/{\gamma_\mathrm{x}}^2)} + A \mathrm{e}^{-b\omega} + c   \label{eq: PL_fitting-2} 
\end{equation}
where $A_{i}$, $\gamma_{i}$, $\omega_{i}$ with $i=\mathrm{a}, \mathrm{x}$ indicates the amplitude, FWHM and resonance of the PNR or X-symmetry-point radiation, respectively. b is the spectral decay constant of the background PL signal. The resonance wavelength (energy) of the PNR obtained from the fit, indicated by the red solid line in Fig.~\ref{cavity_Q_data}a, is at \SI{647}{\nano\meter} (\SI{1.91}{\electronvolt}) which is slightly lower than the documented bare resonance wavelength of the Qdots with \SI{650}{\nano\meter}. The FWHM of the PNR resonance is \SI{0.095}{\electronvolt}, corresponding to a typical Q factor of \si{20}.

\begin{figure*}[htp!]
   \centering
    \includegraphics[width=.99\linewidth]{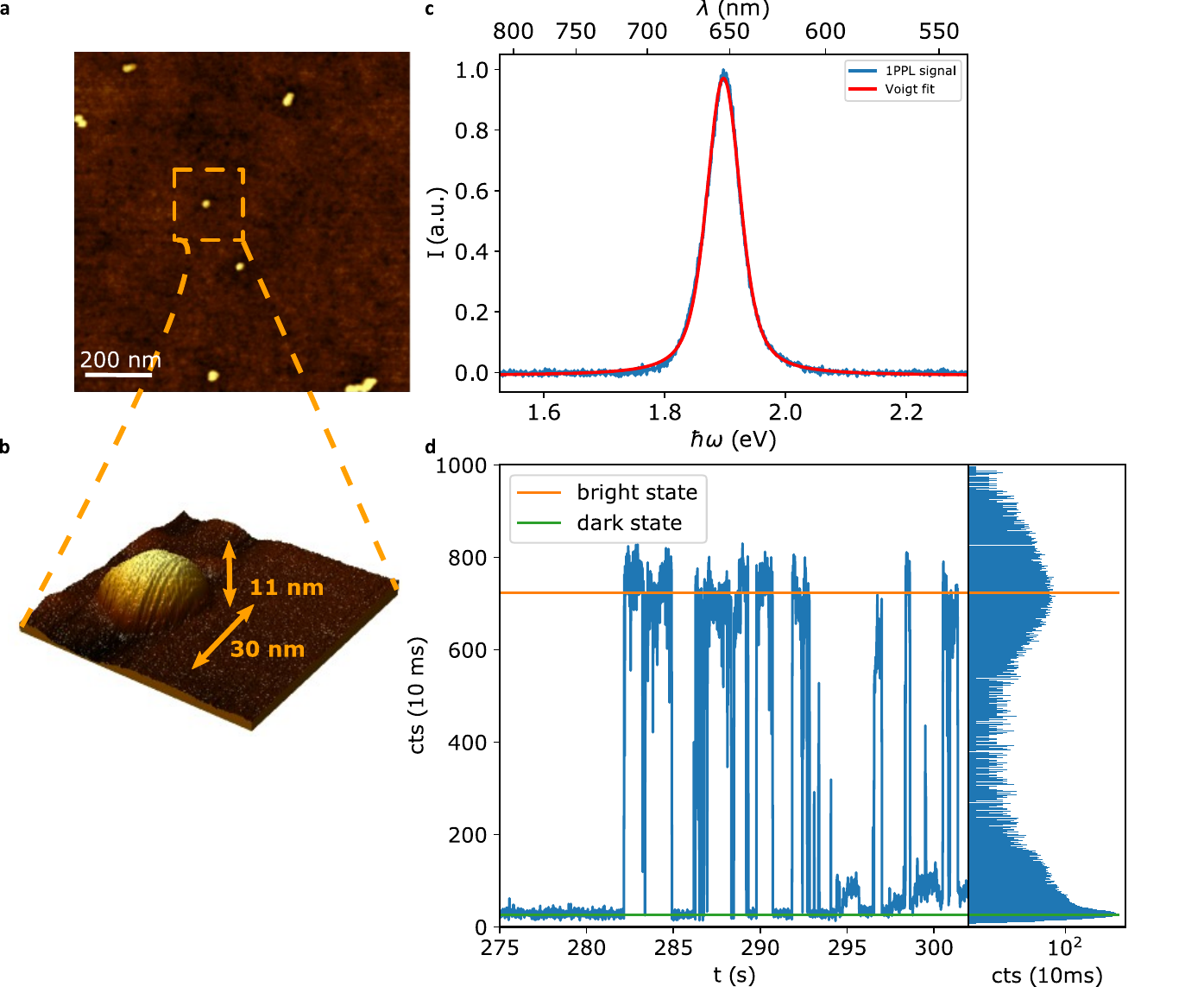}
    \caption{\textbf{a} AFM-scan of a randomly picked area on a glass substrate covered with quantum dots without PMMA layer. Single Qdots as well as smaller agglomerates of nanocrystals (clusters) are distributed randomly after spin-coating. For measurements of strong coupling, the separation between single particles is decisive to perform strong-coupling experiments with single Qdots. A 3D representation of a single Qdot is depicted from the AFM-image in \textbf{b}. It shows the detailed shape and dimensions of an exemplary Qdot. \textbf{c} provides an exemplary PL-spectrum of a single Qdot embedded in a PMMA layer on a glass substrate. The Qdot exhibits a stable peak position at around \SI{655}{\nano\meter}  
In \textbf{d}, a typical recorded time-trace of the Qdot's emitted signal is depicted, providing statistical information about the blinking characteristics of the Qdot (see histogram). For better visualization, just a small excerpt of a longer time-trace is displayed. The histogram on the right indicates a two-state blinking behaviour (bright- and dark state).}
    \label{fig4_QDots}
\end{figure*}


\subsubsection{Quantum dot}
\label{SI:quantum dot}
For all measurements, we use commercial colloidal semiconductor core-shell quantum dots (Qdot 655 ITK Carboxyl Quantum Dots Q21321MP, Thermo Fisher Scientific Inc.) as quantum emitters. The quantum dots consists of a CdSe core and a ZnS shell \cite{brus_quantum_1991}. The shell to some degree suppresses light-induced spectral diffusion and photo-bleaching. To fabricate samples in which single quantum dots are sufficiently separated, an aqueous solution of CdSe/ZnS Qdots is first spin-coated on cleaned microscope coverslips (Gerhard Menzel GmbH). In a second step, a thin (\SI{10}{\nano\meter}) \SI{0.4}{\percent} PMMA film is spin-coated to prevent the quantum dots from being picked-up or being pushed by the scanning PNR-probe. In a typical sample, a lateral separation of at least \SI{1}{\micro\meter} between different Qdots is obtained which guarantees that only one single quantum dot couples to the PNR probe at a time. More details regarding the Qdot-sample fabrication can be found in \cite{gros_near-field_2018}. \\
To characterize Qdot samples, we use a combination of atomic force microscopy (AFM) and confocal microscopy. In Fig.~\ref{fig4_QDots}a, an AFM-scan is displayed showing the typical lateral distribution of single quantum dots and smaller agglomerates on a clean glass substrate prior to PMMA-coating. A 3D representation of a single quantum dot, marked in the 2D AFM-scan, is displayed in Fig.~\ref{fig4_QDots}b. 
Photoluminescence measurements of single Qdots at room temperature excited at a wavelength of \SI{532}{\nano\meter}  show a narrow emission peak at around \SI{655}{\nano\meter}  (see Fig.~\ref{fig4_QDots}c). The emission spectrum of the nanocrystals is stable in time showing no significant spectral fluctuations. A typical time-trace of a quantum dot's PL intensity during permanent excitation recorded by two APDs is presented in Fig.~\ref{fig4_QDots}d. The time trace exhibits typical blinking behavior with a dark and a bright state, as well as the corresponding intensity histogram. 

\begin{figure*}[hbp!]
   \centering
    \includegraphics[width=.7\linewidth]{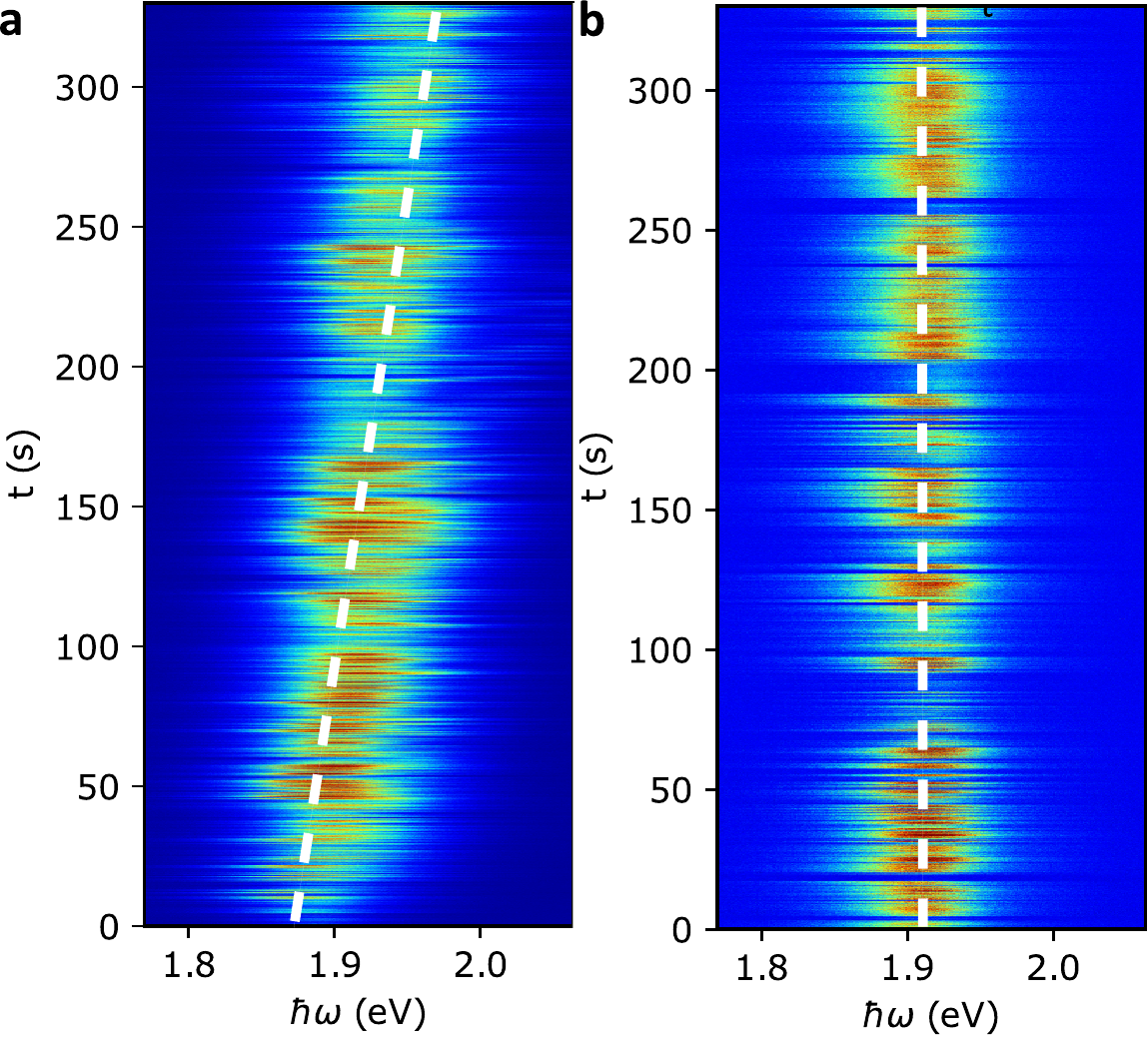}
    \caption{Time traces of single quantum dot Pl spectra at ambient conditions (\textbf{a}) and in argon atmosphere (\textbf{b}). The white-dashed lines serve as guides for the eye indicating the trends of Qdot resonance shifts as a function of time.}
    \label{fig5_QDots}
\end{figure*}

\subsection{Qdot blue-shift}
\label{SI:QD blue-shift}
A light-induced oxygen-dependent blue-shift of colloidal CdSe/ZnSe Qdots was already documented in 1996 by \cite{nirmal_fluorescence_1996}. 
To confirm the presence of such a light-induced oxygen-dependent blue shift for our Qdots we show the PL spectrum of an exemplary individual Qdot recorded over a time span of several minutes during which a spectrum has been recorded every \SI{33}{\milli\second} to trace the spectral changes. The resulting spectra are plotted in Fig.~\ref{fig5_QDots}a as a map with time increasing  from bottom to top and a color-coded PL intensity. A significant blue-shift of about \SI{30}{\nano\meter} can be observed over time in agreement with earlier observations \cite{van_sark_blueing_2002}. To further proof the oxygen dependence we perform the same measurement in an argon atmosphere (see Fig.~\ref{fig5_QDots}b). In absence of oxygen, the Qdot resonance does not shift over time thus confirming the oxygen dependence of the process. For a better visualization we fitted all spectra in Fig.~\ref{fig5_QDots}a and b with an Lorentzian function and plot the peak positions in Fig.~\ref{fig_QD-blue-shift}b. A linear blue shift is observed.\\

A possible reason for the observed blue shift might be oxygen diffusion in the Qdot shell which subsequently oxidizes the Qdot core. The resulting smaller effective core radius then leads to a higher transition energy~\cite{nirmal_fluorescence_1996, van_sark_blueing_2002}. The effect is illustrated in Fig.~\ref{fig_QD-blue-shift}a. We find that the light-induced oxygen-dependent blue-shift effect displayed in Fig.~\ref{fig5_QDots}a continues to increase linearly and can be monitored until the quantum dot is photo-bleached.



\subsection{Classical model simulations}
\label{SI:classical_model}

FDTD simulations were conducted to determine the coupling states between a plasmonic nanorod (PNR) and a single quantum emitter. The dielectric function of single-crystalline gold was obtained from Olmon et al \cite{olmon_dielectric-gold_2012}. To account for the quantum dot exciton state, a Lorentzian function was used to describe its permittivity: $\varepsilon_{\mathrm{Qdot}}(\omega)=\varepsilon_{\infty}+f \omega_0^2/(\omega_0^2-\omega^2-i\gamma_0\omega)$ \cite{schlather2013near}. Here, $\varepsilon_{\infty}$ is the high-frequency component of the CdSe/ZnS quantum dot matrix dielectric function, with a value of \si{5}. The oscillator strength was set to \SI{0.3} and the lowest transition between the exciton state and ground state was at \SI{1.89}{\electronvolt}. The linewidth of the exciton state was \SI{57}{\milli\electronvolt}. The PNR geometry was optimized based on the SEM image, including the rounded corners and edges. By fitting the far-field scattering spectrum with a Lorentzian function, the resonance of the PNR was found to be \SI{1.91}{\electronvolt}, with a cavity decay rate of \SI{95}{\milli\electronvolt}.

\begin{figure*}[htp!]
	   \centering
	    \includegraphics[width=.8\linewidth]{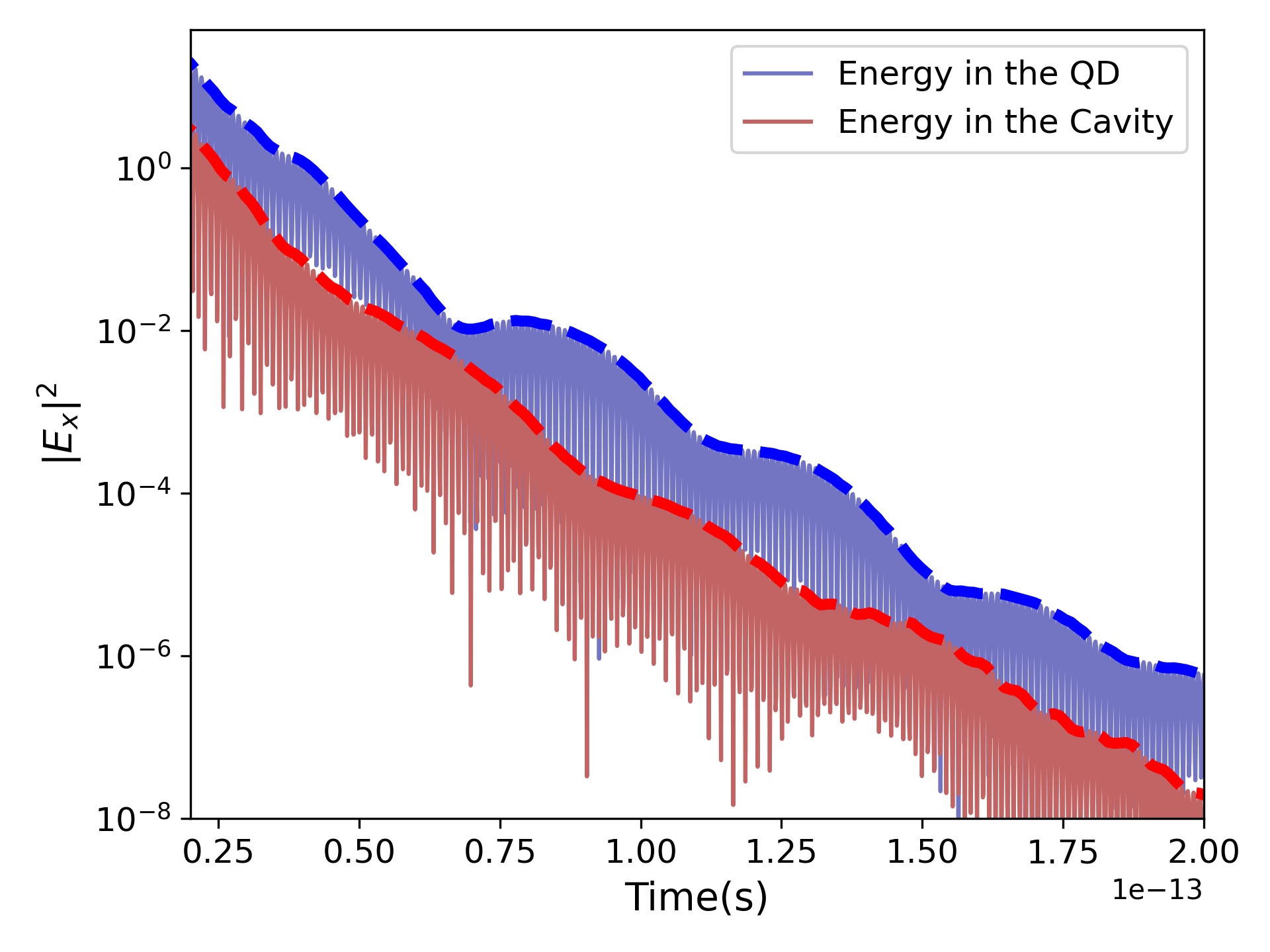}
	    \caption{Evolution of time domain signals inside the slit cavity and quantum dot respectively. Envelopes (blue and red dash lines) indicates the energy exchanges between two system.}
	    \label{figSI-time}
\end{figure*}
\subsubsection{Time-domain simulations}

A characteristic behaviour of strongly coupled systems, such as two coupled oscillators, is their exchange of energy in the time domain after one emitter is excited. For example, in the case of two strongly-coupled gold nanorods with a certain gap, bonding and anti-bonding modes occur in the spectra, but energy exchange also occurs in the time domain. To observe such exchange of energy can serve as the first checkpoint in simulations to confirm that the quantum dot and PNR are indeed strongly coupled. To simulate this, the quantum dot's with its Lorentzian permittivity is placed close to the PNR, and the system is excited by an $x$-polarized plane wave with a pulse duration of about \SI{18}{\femto\second}. Temporal dynamics are recorded from two monitors placed inside the quantum dot and near the PNR, and the $x$-component of the electrical field ${\lvert E_{x} \rvert}^{2}$ is filtered to eliminate irrelevant signals. After the direct influence of the excitation pulse has vanished (at approximately \SI{50}{\femto\second}), the energy exchange pattern indeed starts to appear in the time domain signals. Due to the intrinsic losses in this coupled system, as expected, the energy exchange is visible only for a few oscillations. From the period of this oscillation (approximately \SI{43}{\femto\second}), a Rabi energy of $\sim$\SI{98}{\milli\electronvolt} can be extracted, which is close to the splitting energy observed in our experiments. This kind of time domain simulation is very helpful for validating the model and for obtaining a rough classical picture of strong coupling.
\begin{figure*}[htp!]
	   \centering
	    \includegraphics[width=.99\linewidth]{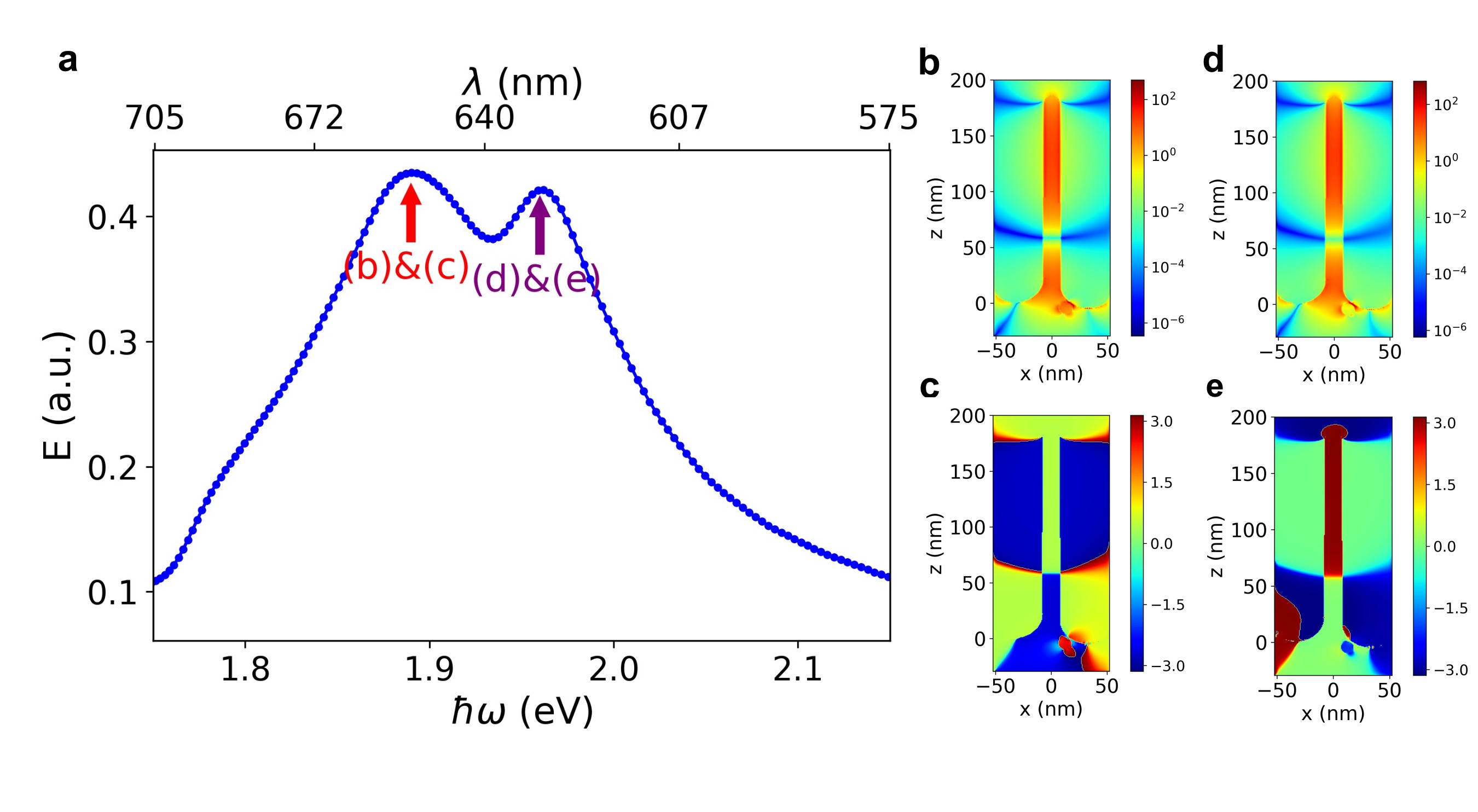}
	    \caption{\textbf{a} Spilt spectra obtained from the FDTD simulations including the quantum dot with a Lorentzian model permittivity. Electric field intensity (\textbf{b} and \textbf{d}) and phase (\textbf{c} and \textbf{e}) at two peaks are extracted indicating two different modes are formed.}
	    \label{fig_SI_mode_fdtd}
	\end{figure*}

\subsubsection{Frequency-domain simulations}
Another feature of strongly coupled systems is the typical splitting in the frequency domain. Using classical simulations, of course only the linear behaviour of the system can be obtained  - the single-photon nonlinearity can only be obtained via a full quantum model. By using a Lorentzian model for the Qdot permittivity to mimic the excitation state the coupling spectrum can be extracted by detecting the electrical field amplitude in the center of the PNR.  It is critical, though, to consider suitable temporal apodization to suppress the contribution of the excitation pulse. Our simulation results are presented in Fig.~\ref{fig_SI_mode_fdtd}(a). Two new peaks ($\omega_-$ \SI{1.88}{\electronvolt},$\omega_+$ \SI{1.96}{\electronvolt}) with a splitting of about \SI{80}{\milli\electronvolt} occur when the resonance of both PNR and quantum dot (($\omega_{\mathrm{PNR}}$,$\omega_{\mathrm{QD}}$)) are set to \SI{1.9}{\electronvolt}. Additionally, mode profiles recorded at the central frequency of these two peaks ($\omega_{\pm}$) can be obtained and the intensity and phase of the respective near-fields ${\lvert E_{x} \rvert}$ can be investigated, as displayed in Fig.~\ref{fig_SI_mode_fdtd}(b-d). The near-field profile at $\omega_-$ is similar to the second-order mode of the bare PNR. In contrast, the near-field profile at $\omega_+$ shows large field enhancement in the Qdot region. In addition, a typical phase shift occurs, indicating that a new mode is formed due to strong coupling. The simulated spectrum resembles our experimental spectra for the case of zero detuning between PNR and Qdot.

\subsubsection{Estimation of the coupling strength}
\begin{figure*}[htp!]
   \centering
    \includegraphics[width=.99\linewidth]{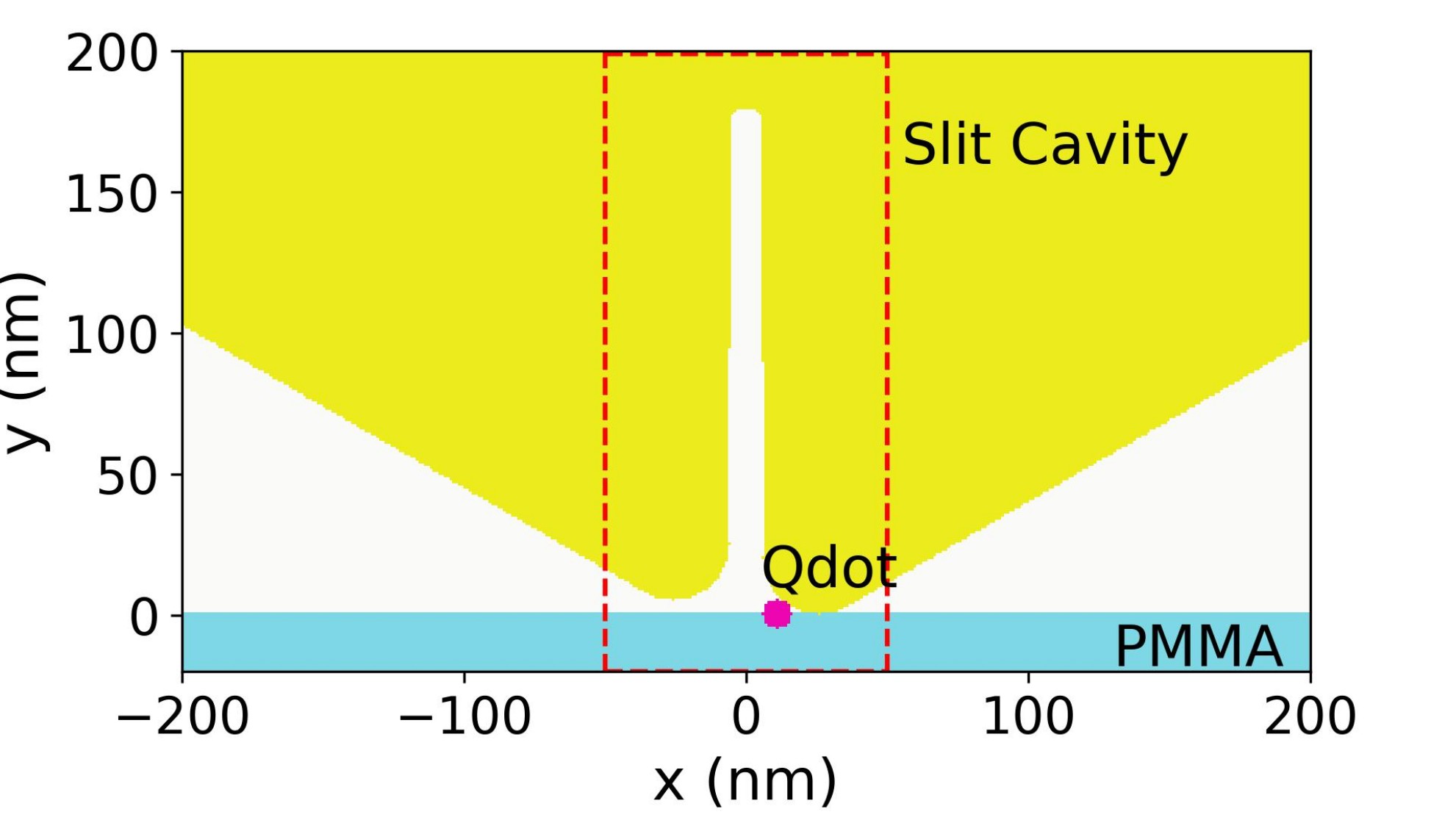}
    \caption{Sketch of the simulation area in FDTD in a 2D projection showing the PNR (yellow), the Qdot (pink) and the glass substrate (bluegreen). The red-dashed rectangle indicates the region of interest in our model.}
    \label{fig_FDTD_simulationarea}
\end{figure*}

The coupling energy between the PNR and quantum dot $\boldsymbol{g}$ is determined by the scalar product of the dipole moment $\boldsymbol{\mu}$ of the quantum dot with the vacuum field amplitude $E_0$ at position $\boldsymbol{r}$. $E_0(\boldsymbol{r})$ is defined as
\begin{equation}
    E_0(\boldsymbol{r})=\sqrt{\frac{\hbar\omega}{2\varepsilon_0V_{\mathrm{eff}}(\boldsymbol{r})}} \label{eq: vacuumfield} 
\end{equation}
where $\hbar\omega$ is the photon energy, $\varepsilon_0$ is the permittivity of free space, and $V_{\mathrm{eff}}(\boldsymbol{r})$ is the effective cavity mode volume of the PNR. Equation~\eqref{eq: vacuumfield} shows that $\boldsymbol{g}$ scales with $\sqrt{1/V_{\mathrm{eff}}}$. One possible way to estimate the coupling energy therefore is to calculate the effective mode volume. However, plasmonic cavities often have low Q factors, causing integral divergence when using the common normal mode prescription. To address this issue, we use quasi-normal modes (QNM) with complex frequencies to determine the effective mode volume, following Sauvan et al.~\cite{sauvan_theory_2013}. The effective mode volume is expressed as
\begin{equation}
    V_{\mathrm{eff}}(\boldsymbol{r})=\frac{\int\left(\vec{E}\cdot\frac{\partial(\omega\varepsilon(\boldsymbol{r}))}{\partial\omega}\vec{E}-\vec{H}\cdot\frac{\partial(\omega\mu(\boldsymbol{r}))}{\partial\omega}\vec{H}\right)\mathrm{d}^3\boldsymbol{r}}{2\varepsilon_0\varepsilon(\boldsymbol{r}){\lvert\vec{E}(\boldsymbol{r})\rvert}^2} \label{eq: effectivemode} 
\end{equation}
where $\omega$ becomes a complex number whose imaginary part is determined from $Q=-\Re(\omega)/2\Im(\omega)$, $\varepsilon(\boldsymbol{r})$ and $\mu(\boldsymbol{r})$ are the relative permittivity and permeability of the simulation area, and $\vec{E}$ and $\vec{H}$ the electric and magnetic field distributions inside the PNR under x-polarized plane wave excitation. The simulation configuration is shown in Fig.~\ref{fig_FDTD_simulationarea}. Apodization was applied according to Ge et al.~\cite{ge_design_2014}. The simulation area is defined by a homogeneous mesh with size of \SI{0.5}{\nano\meter}. \\

To determine the dipole moment, we use the relation between the dipole moment and the oscillator strength of the exciton transition, which is given by $\mu=\sqrt{\frac{f\hbar}{2m_\mathrm{e}\omega}}$, where $m_e$ is the electron mass. In our case, only the lowest optical transition is considered for strong coupling behaviour and its oscillator strength is determined either via lifetime or absorption measurements. Here, we measure the fluorescence lifetime $\tau$ to deduce the oscillator strength by \cite{leistikow_size-dependent_2009, thranhardt_relation_2002}:
\begin{equation}
    f=\frac{6m_\mathrm{e}\epsilon_0\pi c^3}{q^2n\omega^2\tau} \label{eq: oscillator strength} 
\end{equation}
where $m_e$ and q is the electron mass and charge respectively, $\epsilon_0$ is the vacuum permittivity, c is the speed of light in the vacuum, n is the refractive index of the surrounding medium. Fitting the $g^2(\tau)$ measurement yields a lifetime of the quantum dot of \SI{43.8}{\nano\second}. Therefore, a dipole moment of \SI{5} Debye is assigned to estimate the coupling energy.

\subsection{Quantum model of strong coupling}
\label{SI:quantum_model}
\begin{figure*}[htp!]
   \centering
    \includegraphics[width=.99\linewidth]{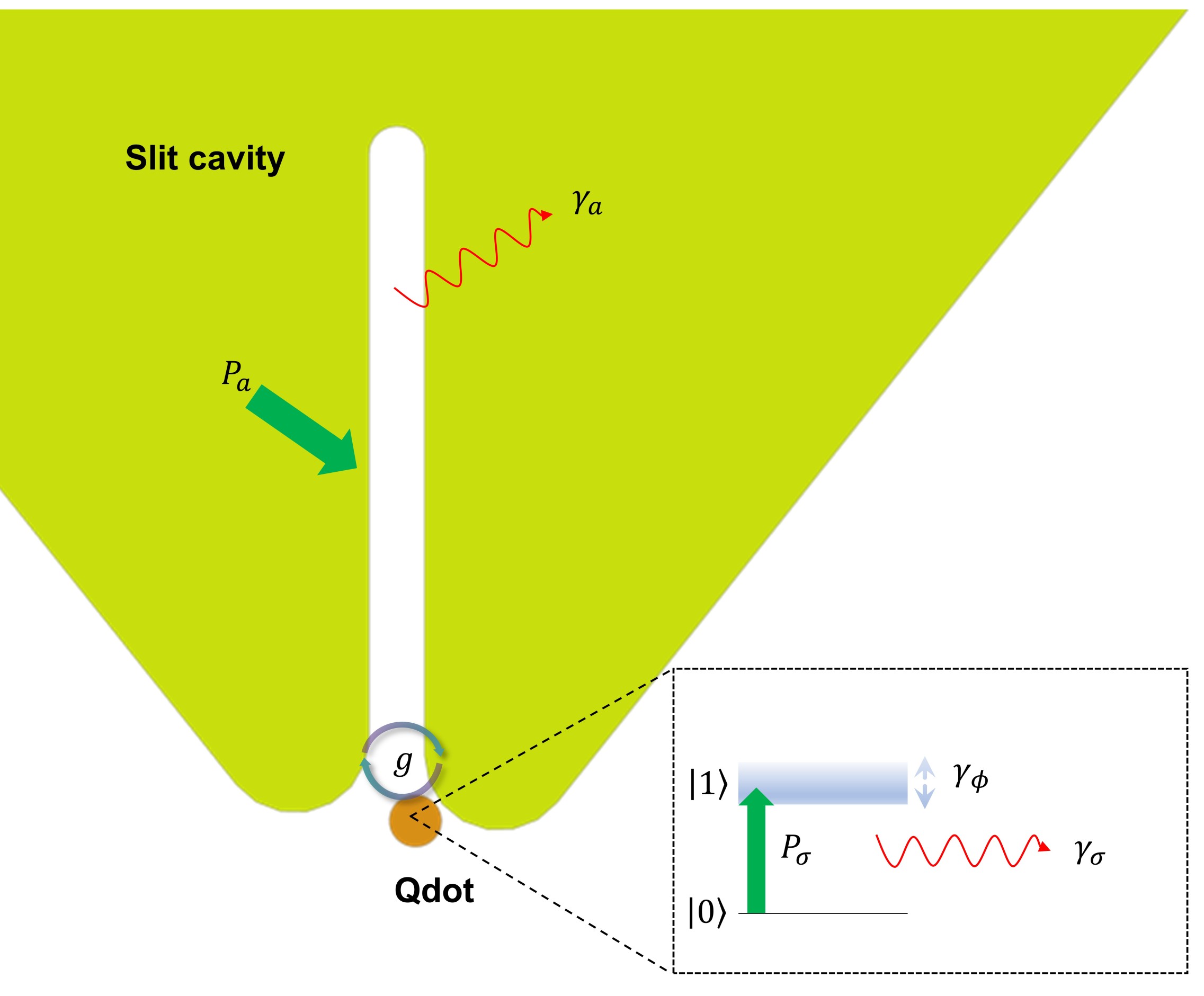}
    \caption{Illustration of the coupling behaviour between slit cavity and Quantum dots. The slit cavity is coupled with a quantum dot (Qdot) by a strength $g$. For the slit cavity, we consider the incoherent pump rate $P_{a}$ and dissipative rate $\gamma_{a}$. A two-level system is used to mimic the dynamics of Qdot as (inset). Similarly, the incoherent pump rate $P_{\sigma}$ and radiavtive loss rate $\gamma_{\sigma}$ are accounted for the model. Also, the dephasing rate $\gamma_{\phi}$ is included.}
    \label{fig5_ill}
\end{figure*}

Light-matter strong coupling can be described by the Jaynes-Cummings Hamiltonian ($\hbar=1$) \cite{bruce_jaynes-cummings_1993}:
\begin{equation}
    H=\omega_a a^{\dagger}\mathrm{a}+\omega_{\sigma} \sigma^{\dagger}\sigma +g(a^{\dagger}\sigma+a\sigma^{\dagger}) \label{:JCmodel}
\end{equation}
where $a$ $(a^{\dagger})$ denote annihilation (creation) operators of a single Bosonic mode with the energy of $\omega_a$, $\sigma^{\dagger}$ $(\sigma)$ represents the raising (lowering) operators of a two-level system with transition energy $\omega_{\sigma}$. $\sigma_{x,y,z}$ are Pauli matrices. These two systems (the cavity and the quantum dot) are coupled with a coupling strength $g$. To obtain the spectrum of $H$, the dynamics of this hybrid system are studied by solving the master equation:
\begin{equation}
    \frac{\mathrm{d}}{\mathrm{d}t}\rho=-i[H,\rho]+\sum\limits_{i}\mathcal{L}_i(\rho)   \label{:masterequation}
\end{equation}
where $\rho$ denotes the density matrix, and $\mathcal{L}_i$ the Lindblad superoperators accounting for all kinds of dissipative contributions of the dynamics, including radiative decay ($\gamma_{a,\sigma}$), incoherent pumping ($P_{a,\sigma}$), and pure dephasing ($\gamma_{\phi}$). The sketch in Fig.~\ref{fig5_ill} illustrates the detailed physical meaning of each parameter. The Lindblad operators have the following form \cite{delValle_luminescence_2009, slowik_strong-coupling_2013}:
\begin{equation}
    \begin{split} 
    \sum\limits_{i}\mathcal{L}_i(\rho) & = \sum\limits_{c=\mathrm{a},\sigma}\frac{\gamma_\mathrm{c}}{2}\left(2c\rho c^{\dagger}-c^{\dagger}c\rho-\rho c^{\dagger}c\right) +\sum\limits_{c=\mathrm{a},\sigma}\frac{P_\mathrm{c}}{2}\left(2c\rho c^{\dagger}-c^{\dagger}c\rho-\rho c^{\dagger}c\right) \\
    &+\frac{\gamma_{\phi}}{2}\left(\sigma^{\dagger}\sigma\rho\sigma^{\dagger}\sigma-\sigma^{\dagger}\sigma\rho-\rho\sigma^{\dagger}\sigma\right)
    \label{:lindbladoperators}
    \end{split}
\end{equation}

Here, we assume the cavity emission is dominant in the hybrid system and the direct emission of the quantum dot is neglected \cite{delValle_quantum_2009}.Under theses assumptions, the emission spectrum $S(\omega)$ is obtained as: 
\begin{equation}
    S(\omega)\propto\Re\left(\int_{0}^{\infty}\left\langle a^{\dagger}(\tau)a(0)\right\rangle e^{-i\omega\tau}\mathrm{d}\tau\right)\; . 
    \label{:spectra}
\end{equation}
The numerical simulations are performed by using the Python module Quantum Box (Qutip)~\cite{johansson_qutip_2013}. The parameters used for generating the anticrossing map Fig.~\ref{fig3_anticrossing}b are $g$=\SI{50}{\milli\electronvolt}, $\omega_\mathrm{a}$=\SI{1.91}{\electronvolt}, $\gamma_\mathrm{a}$=\SI{95}{\milli\electronvolt}, $P_\mathrm{a}$=\SI{0.16}{\milli\electronvolt}, $\gamma_\sigma$=\SI{15.0}{\nano\electronvolt}, $\gamma_\phi$=\SI{55.7}{\milli\electronvolt},$P_\sigma$=\SI{0.16}{\milli\electronvolt}. The data in the map is normalized row by row.

 \label{SI:spectrum_fitting}
In our experiments, we utilize a green laser to incoherently excite the quantum dot, hence only the incoherent pumping of the quantum dot $P_{\sigma}$ term should be taken into consideration during fitting. Nevertheless, some studies have shown that the cavity pumping has non-negligible photonic contributions to the overall spectrum \cite{delValle_quantum_2009}. Therefore, we consider both of these two incoherent pumping terms in our fitting. The spectra displayed in Fig.~\ref{fig3_anticrossing}a are processed using the fast Fourier transform algorithm (FFT) to smooth out high-frequency noise. Subsequently, the spectra are normalized between \SI{0} and \SI{1}. 
The fitting approach is such that only the quantum dot resonance $\omega_{\sigma}$ and the cavity (quantum dot) incoherent pumping rates $P_{\mathrm{a},\sigma}$ are treated as free parameters, whereas the coupling strength $g$ is optimized but kept constant for all the spectra. The cavity resonance $\omega_{\mathrm{a}}$ (\SI{1.91}{\electronvolt}) and the cavity decay rate $\gamma_{\mathrm{a}}$ (\SI{95}{\milli\electronvolt}) are also kept fixed. They are extracted from the PL spectrum of the bare PNR. It is important to correctly describe the incoherent processes affecting the QDot. Typically, the total decay rate $\gamma_{\mathrm{QD}}$, which is obtained from the experimentally observed linewidth of the bare Qdot emission, is composed of the radiative decay described by $\gamma_{\sigma}$ and pure dephasing rate, $\gamma_{\phi}$, as $\gamma_{QD}=\gamma_{\sigma}/2+\gamma_{\phi}$. The radiative decay rate $\gamma_{\sigma}$ (\SI{15.0}{\nano\electronvolt}) is extracted from the radiative lifetime $\tau$ (\SI{43.8}{\nano\second}) of the bare Qdot using $\gamma_{\sigma}\tau=\hbar$. It is evident that the pure dephasing mechanism is the dominant mechanism that contributes to the linewidth of a QDot. Our previous study shows that the Qdot linewidth would be slightly affected by the presence of the gold tip \cite{gros_near-field_2018}. Typically, the fitting process makes use of non-linear least squares minimization. However, some spectra still appear noisy even after smoothing, such that the parameters needed to be adjusted to match the experimental results. The fitting parameters can be found in the table below.

\begin{table}[h!]
\begin{tabular}{cccccccc}
\hline
ID       & \textbf{\begin{tabular}[c]{@{}c@{}} $g$ ({\si{\milli\electronvolt}})\end{tabular}} & \textbf{\begin{tabular}[c]{@{}c@{}} $\omega_{\sigma}$ ({\si{\electronvolt}})\end{tabular}} & \textbf{\begin{tabular}[c]{@{}c@{}} $\gamma_a$ ({\si{\milli\electronvolt}})\end{tabular}} & \textbf{\begin{tabular}[c]{@{}c@{}} $P_a$ ({\si{\milli\electronvolt}})\end{tabular}} & \textbf{\begin{tabular}[c]{@{}c@{}} $\gamma_{\sigma}$ ({\si{\nano\electronvolt}})\end{tabular}} & \textbf{\begin{tabular}[c]{@{}c@{}} $P_{\sigma}$ ({\si{\milli\electronvolt}})\end{tabular}} & \textbf{\begin{tabular}[c]{@{}c@{}} $\gamma_{\phi}$ ({\si{\milli\electronvolt}})\end{tabular}} \\ \hline
Spec \#1 & 50.0                                                                     & 1.870                                                                                    & 95.5                                                                               & 13.37                                                                        & 15.0                                                                                  & 4.90                                                                             & 55.7                                                                                    \\
Spec \#2 & 50.0                                                                     & 1.895                                                                                    & 95.5                                                                               & 0.16                                                                         & 15.0                                                                                  & 0.16                                                                             & 55.7                                                                                    \\
Spec \#3 & 50.0                                                                     & 1.925                                                                                    & 95.5                                                                               & 0.16                                                                         & 15.0                                                                                  & 0.16                                                                             & 55.7                                                                                    \\
Spec \#4 & 50.0                                                                     & 1.950                                                                                    & 95.5                                                                               & 7.95                                                                         & 15.0                                                                                  & 7.95                                                                             & 50.9                                                                                    \\
Spec \#5 & 50.0                                                                     & 1.970                                                                                    & 95.5                                                                               & 2.71                                                                         & 15.0                                                                                  & 2.70                                                                             & 49.2                                                                                    \\
Spec \#6 & 50.0                                                                     & 1.990                                                                                    & 95.5                                                                               & 0.95                                                                         & 15.0                                                                                  & 0.95                                                                             & 47.7                                                                                    \\
Spec \#7 & 50.0                                                                     & 2.040                                                                                    & 95.5                                                                               & 2.00                                                                         & 15.0                                                                                  & 2.00                                                                             & 55.7                                                                                    \\ \hline
\end{tabular}
\end{table}

\subsection{Selection of spectra}

\begin{figure*}[htp!]
   \centering
    \includegraphics[width=.99\linewidth]{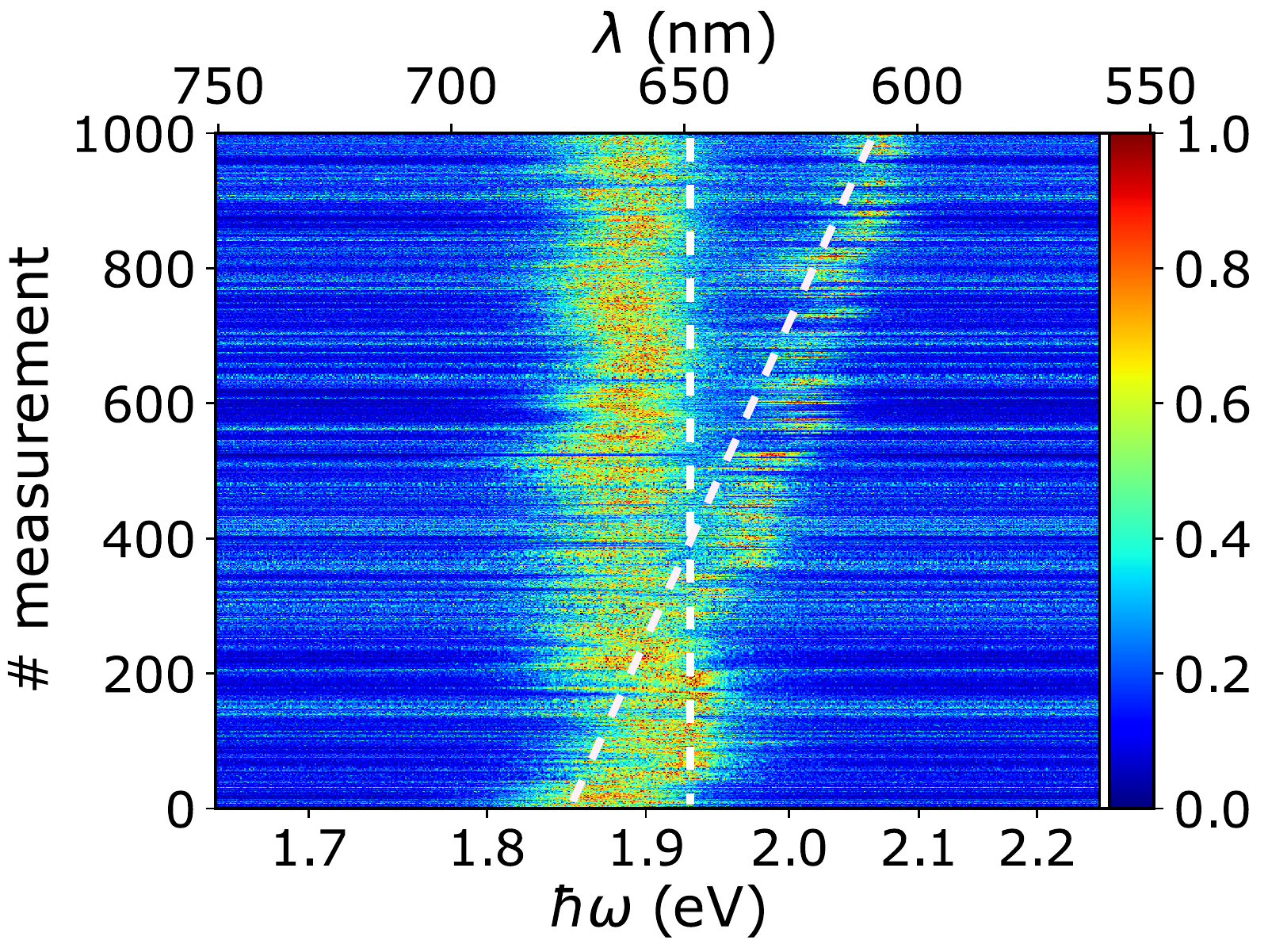}
    \caption{PL spectra arranged by time sequence. The split peaks (orange dash line) and slit cavity resonance (white dash line) are labeled.}
    \label{fig_spectra_selection}
\end{figure*}

In the main text, the anti-crossing spectra presented in Fig.~\ref{fig3_anticrossing}(a) were obtained from the  sequential measurement based on their time sequence neglecting spectra according to certain criteria (see below). A complete coupling spectra map is displayed in Fig.~\ref{fig_spectra_selection}. Each horizontal line in the plot represents one measurement with an integration time of \SI{33}{\milli\second}. The two branches in the map illustrate how the coupling spectra evolve along the time sequence. The left branch starts from the lower energy and gradually gets closer to the cavity resonance as the initial resonance of Qdot is slightly red-detuned. The other branch shifts away to the higher energy due to the blue-detuned Qdot (caused by light-induced oxidation).

Dark curves, showing no peaks, were removed, as the Qdot occasionally went into the 'off' state. Therefore, the y-axis in this case does not indicate equivalent time spacing, but still strictly follows the time trend.  

In general, all the spectra can be fitted by the quantum model, and the corresponding Qdot resonance can be extracted. In Fig.~\ref{fig3_anticrossing}a we show only 7 spectra to demonstrate the typical anti-crossing behavior with roughly equal energy spacing of Qdot resonance.

\bibliography{lit}

\end{document}